\documentclass[fleqn,usenatbib]{mnras}

\usepackage{newtxtext,newtxmath}
\usepackage{comment}
\usepackage{longtable}
\usepackage{tablefootnote}
\usepackage{threeparttablex}
\usepackage{graphicx}
\usepackage{txfonts}
\usepackage{verbatim}
\usepackage{hyperref}
\usepackage{tabularx}
\usepackage{array}
\usepackage{adjustbox}
\usepackage{pdflscape}
\usepackage[dvipsnames]{xcolor}
\newcommand{\lsun}{$\log$L/$L_{\odot}\,$}
\newcommand{\msun}{$M$/$M_{\odot}\,$}

\usepackage{multicol}
\usepackage{graphicx}
\usepackage{natbib}
\defcitealias{Desomma2020a}{DS20a}
\defcitealias{Desomma2020b}{DS20b}
\defcitealias{Caputo2000}{Caputo2000}
\defcitealias{Marconi2005}{Marconi2005}
\defcitealias{Marconi2010}{Marconi2010}

\usepackage[bottom]{footmisc}
\graphicspath{{./}{figures/}}
\usepackage[T1]{fontenc}

\DeclareRobustCommand{\VAN}[3]{#2}
\let\VANthebibliography\thebibliography
\def\thebibliography{\DeclareRobustCommand{\VAN}[3]{##3}\VANthebibliography}

\usepackage{graphicx}	
\usepackage{amsmath}	
\usepackage{amssymb}	

\title[The Cepheid Period-Age-metallicity and Period-Age-Color-Metallicity relations]{Period-Age-Metallicity and Period-Age-Color-Metallicity relations for Classical Cepheids: an application to the \textit{\textbf{Gaia}} EDR3 sample}

\author[G. De Somma et al.]{
Giulia De Somma,$^{1}$$^{2}$\thanks{E-mail: giulia.desomma@inaf.it}
Marcella Marconi,$^{1}$
Santi Cassisi,$^{3}$$^{4}$ 
Vincenzo Ripepi,$^{1}$
\newauthor{Adriano Pietrinferni},$^{3}$
Roberto Molinaro,$^{1}$
Silvio Leccia$^{1}$
and Ilaria Musella$^{1}$
\\
$^{1}$ INAF-Osservatorio Astronomico di Capodimonte, Via Moiariello 16, 80131 Napoli, Italy\\
$^{2}$ Istituto Nazionale di Fisica Nucleare (INFN)-Sez. di Napoli, Compl. Univ.di Monte S. Angelo, Edificio G, Via Cinthia, 80126 Napoli, Italy\\
$^{3}$ INAF-Osservatorio Astronomico d'Abruzzo, Via Maggini sn, 64100 Teramo, Italy\\
$^{4}$ Istituto Nazionale di Fisica Nucleare (INFN) - Sezione di Pisa, Universit\'a di Pisa, Largo Pontecorvo 3, 56127 Pisa, Italy
}

\date{Accepted 2021 September 10. Received 2021 September 10; in original form 2021 July 01}

\pubyear{2021}

\begin{document}
\label{firstpage}
\pagerange{\pageref{firstpage}--\pageref{lastpage}}
\maketitle

\begin{abstract}
Based on updated pulsation models for Classical Cepheids, computed for various assumptions about the metallicity and helium abundance, roughly representative of pulsators in the Small Magellanic Cloud ($Z$=$0.004$ and $Y$=$0.25$), Large Magellanic Cloud ($Z$=$0.008$ and $Y$=$0.25$), and M31 ($Z$=$0.03$ and $Y$=$0.28$), and self-consistent updated evolutionary predictions, we derived Period-Age and multi-band Period-Age-Color relations that also take into account variations in the Mass-Luminosity relation. These results, combined with those previously derived for Galactic Cepheids, were used to investigate the metallicity effect when using these variables as age indicators. In particular, we found that a variation in the metal abundance affects both the slope and the zero point of the above-mentioned relations. The new relations were applied to a sample of Gaia Early Data Release 3 Classical Cepheids. The retrieved distribution of the individual ages confirms that a brighter Mass-Luminosity relation produces older ages and that First Overtone pulsators are found to be concentrated towards older ages with respect to the Fundamental ones at a fixed Mass-Luminosity relation. Moreover, the inclusion of a metallicity term in the Period-Age and Period-Age-Color relations slightly modifies the predicted ages. In particular, the age distribution of the selected sample of Galactic Cepheids is found to be shifted towards slightly older values, when the F-mode canonical relations are considered, with respect to the case at a fixed solar chemical composition. A marginally opposite dependence can be found in the noncanonical F-mode and canonical FO-mode cases.

\end{abstract}




\section{Introduction}

Classical Cepheid (CC) Period-Luminosity (PL) and Period-Luminosity-Color (PLC) relations are the basis for the absolute calibration of the extragalactic distance scale, which leads to a local evaluation of the Hubble constant ($H_0$) \citep[see e.g.][]{Freedman2001, Riess2018, Riess2021, Saha2006}. This role of CCs has known a renewed interest in the last few years due to the debate on the so-called $H_0$ tension, the discrepancy between early Universe measurements of $H_0$ based on the Cosmic Microwave background analysis, and late Universe estimates of the constant, based on the extragalactic distance scale \citep[see e.g.][and references therein]{Verde2019}.  

In addition, CCs are also known to obey Period-Age (PA) and Period-Age-Color (PAC) relations \citep[see e.g.][and references therein]{Anderson2016, Bono2005,Efremov2003}. Indeed, the combination of the period-mean density relation and the Stefan-Boltzmann Law leads to a period-luminosity-mass-effective temperature relation that, once it assumes a mass-luminosity relation and a transformation into the observational plane, leads to a period-luminosity-color relation. As the luminosity is related to the mass and the evolving mass is anticorrelated with the stellar age, a period-age-color relation arises. The period-age relation is then found by averaging the period over the color extension of the instability strip for any given age. Therefore, this class of pulsating stars represents a powerful tool for measuring individual stellar ages, and hence, inferring tight constraints on the star formation history of the stellar populations to which they belong.

A PA relation for CCs can be obtained via a semi-empirical approach combining the measurement of the pulsational properties of a sample of CCs in a star cluster, with the corresponding age estimate obtained via the isochrone fitting technique applied to the cluster Color-Magnitude Diagram (CMD) \citep[see, for a recent analysis,][and references therein]{Medina21, Zhou2021}.

A completely different approach \citep{Bono2005} relies on a theoretical scenario in which the Mass-Luminosity (ML) relation and evolutionary lifetimes during the core He-burning stage, as predicted by stellar models \citep[see e.g.][and references therein]{BCM2000b, Chiosi1993}, are connected with the pulsation relation (linking the period with the stellar mass, luminosity and effective temperature) obtained via suitable pulsational models so as to derive the PA relation. The use of the PAC relation with respect the PA one represents an improvement, applied in order to overcome the problem related to the finite color width of the instability strip (IS) (\citet[][]{Caputo2000, Marconi2005, Marconi2010} and \citet[][hereafter DS20a]{Desomma2020a}).

In a recent work \citep[][hereafter DS20b]{Desomma2020b}, we computed updated PA and PAC relations - in the Gaia DR2 photometric passbands \citep{Brown2018, Prusti2016} - for solar metallicity CCs, by combining an extended and homogeneous set of nonlinear convective pulsation models obtained for several assumptions on the ML relation, and the efficiency of superadiabatic convection (\citetalias[we refer to][for more details]{Desomma2020a}), with the updated BaSTI library of stellar evolutionary predictions presented in \cite{hidalgo18}.

The obtained PA and PAC relations were applied to a subset of Gaia Data Release 2 Galactic Cepheids \citepalias[see][for details]{Desomma2020a,Desomma2020b}. The retrieved age distributions confirm that the effect of a variation in the efficiency of superadiabatic convection in the pulsational computations is negligible, whereas older ages are obtained when a brighter mass-luminosity relation, due to a combination of mild overshooting and/or rotation, and/or mass-loss, is adopted. 

The aim of the present investigation is to extend the theoretical calibration of the PA and PAC relations to other chemical compositions, representative of the typical chemical composition of the Small and Large Magellanic Clouds ($Z$=$0.004$ and $Z$=$0.008$ with $Y$=$0.25$), and the Andromeda (M31) galaxy ($Z$=$0.03$ with $Y$=$0.28$).

A complete analysis of the pulsational models corresponding to these chemical abundances, including the investigation of the effects related to changes in the adopted ML relation and efficiency of super-adiabatic convection, will be presented in a companion paper (De Somma et al. 2021, in prep.). Here, we focus our analysis on the derivation of PA and PAC relations in several photometric systems, and of the first Period-Age-Metallicity (PAZ) and Period-Age-Color-Metallicity (PACZ) relationships.

The structure of the paper is as follows: in Section 2, we present the new metal-dependent theoretical scenario for CCs, based on an extended set of evolutionary tracks and nonlinear convective pulsation models; in Section 3, we derive the new PA and multi-filter PAC relations by varying the chemical compositions which are then applied to a subset of Gaia Early Data Release 3 (EDR3) CCs in Section 4; the conclusions and possible future developments of the presented investigation are discussed in Section 5.

\section{A metal dependent theoretical scenario}

In order to retrieve the theoretical PA and PAC calibrations, it is mandatory to
combine the stellar evolutionary framework with the pulsation model predictions:
\begin{itemize}
\item Evolutionary models provide the relevant information about the morphology of the
evolutionary tracks during the core He-burning phase (the so-called blue loop), and the evolutionary lifetimes in this phase;
\item The pulsation scenario gives crucial information about the IS boundaries, and the relation between the pulsation period and stellar evolutionary properties such as current mass, luminosity and effective temperature.
\end{itemize}

In the following subsections, we provide a concise description of the main characteristics of the adopted evolutionary and pulsation frameworks.

\subsection{The evolutionary framework}

The sets of stellar models adopted in the present investigation are fully consistent with those adopted for deriving the PA and PAC relations for a solar chemical composition by \citetalias{Desomma2020b}. These stellar model sets correspond to the updated version of the BaSTI stellar evolutionary library for the solar scaled heavy element distribution, presented by \cite{hidalgo18} and available at the URL:\url{http://basti-iac.oa-abruzzo.inaf.it.}

A complete description of the physical inputs and assumptions adopted for computing these models is provided by \cite{hidalgo18}, while a brief summary of the features more relevant for the investigation of the CC properties can be found in \cite{Desomma2020b}. Here, it is enough to mention that for this analysis we selected stellar models in the $4M_\odot$ and $11M_\odot$ mass range for the following chemical compositions: Z=0.004, Y=0.252; Z=0.008, Y=0.257; and Z=0.03, Y=0.284\footnote{The chemical compositions of the stellar model sets are quite consistent with those adopted for the pulsational computations. The extremely small differences in the initial He abundances have no impact at all on present investigation.}

The BaSTI library provides evolutionary predictions for both stellar models, accounting for a moderate core convective overshooting during the core H-burning stage (noncanonical models), and completely neglecting the occurrence of any physical process able to increase the size of the convective core above the canonical value predicted by the Schwarzschild criterion (canonical scenario) \citep[we refer to][for a detailed discussion on this topic]{hidalgo18}.

\begin{figure*}
\includegraphics[width=0.88\textwidth]{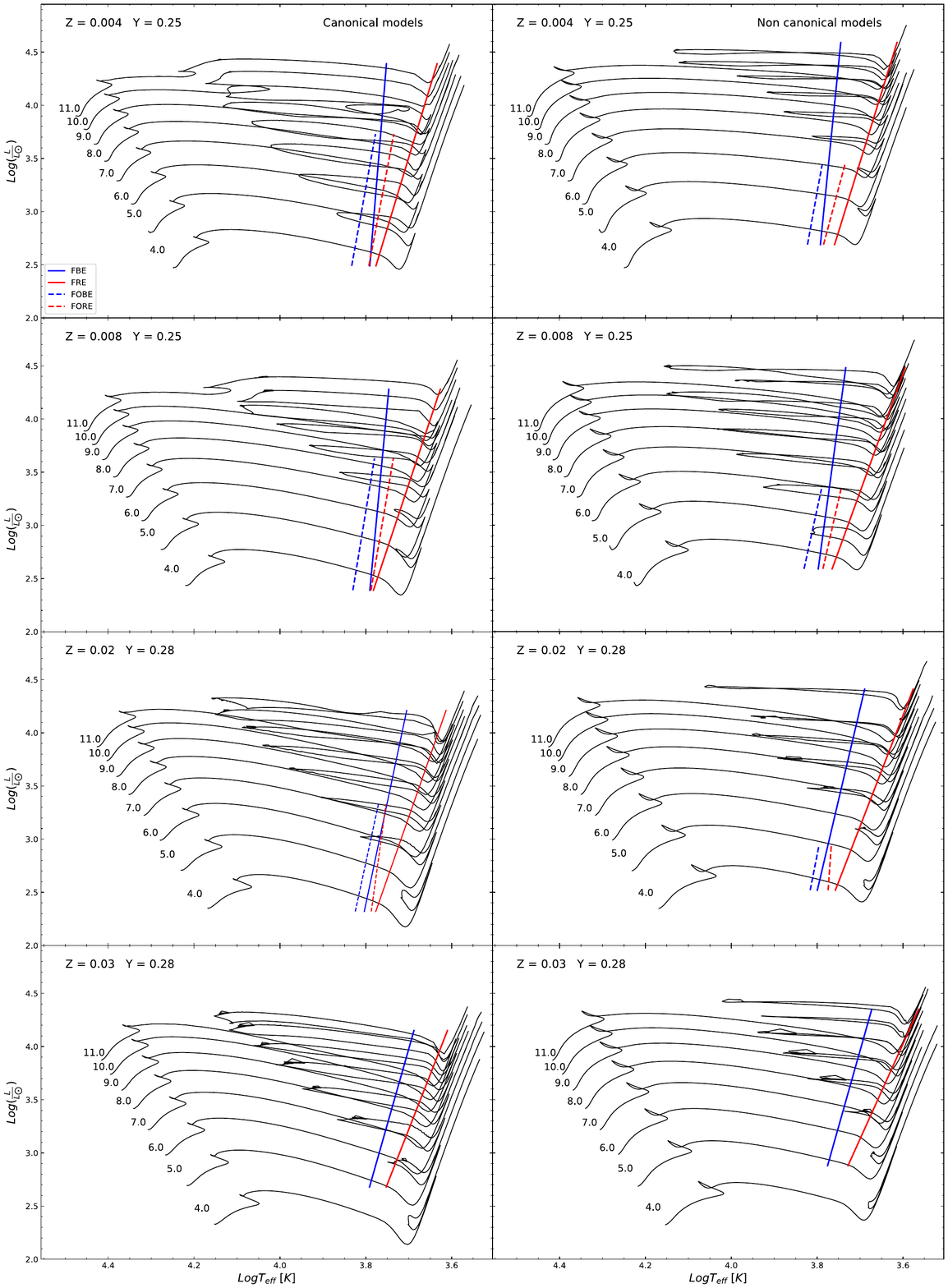}
\caption{\label{fig:strips_tracks_z} The location, in the HR diagram, of a sub-sample of massive and intermediate-mass stellar models from the BaSTI-IAC library, for various assumptions of the initial chemical compositions (see labels). The predicted boundaries of the IS, fundamental blue edge (FBE), fundamental red edge (FRE), first overtone blue edge (FOBE) and first overtone red edge (FORE), for both F (solid line) and FO-mode pulsators (dashed line) are also shown. Left panels refer to the canonical scenario, while the right panels correspond to the noncanonical scenario (see text for more details).}
\end{figure*}

A subsample of the stellar evolutionary tracks adopted in this work is shown in Fig.~\ref{fig:strips_tracks_z}.

In order to allow the derivation of the PAC relation, the selected sets of stellar models have been converted from the theoretical Hertzsprung-Russell (HR) diagram to various photometric systems, such as the Johnson-Cousin, the Gaia EDR3 \citep{Brown2021} and the James Webb Space Telescope (JWST) NIRCam ones. As such, the procedure described in detail in \cite{hidalgo18} has been adopted.

\begin{figure*}
\centering
\includegraphics[width=\textwidth]{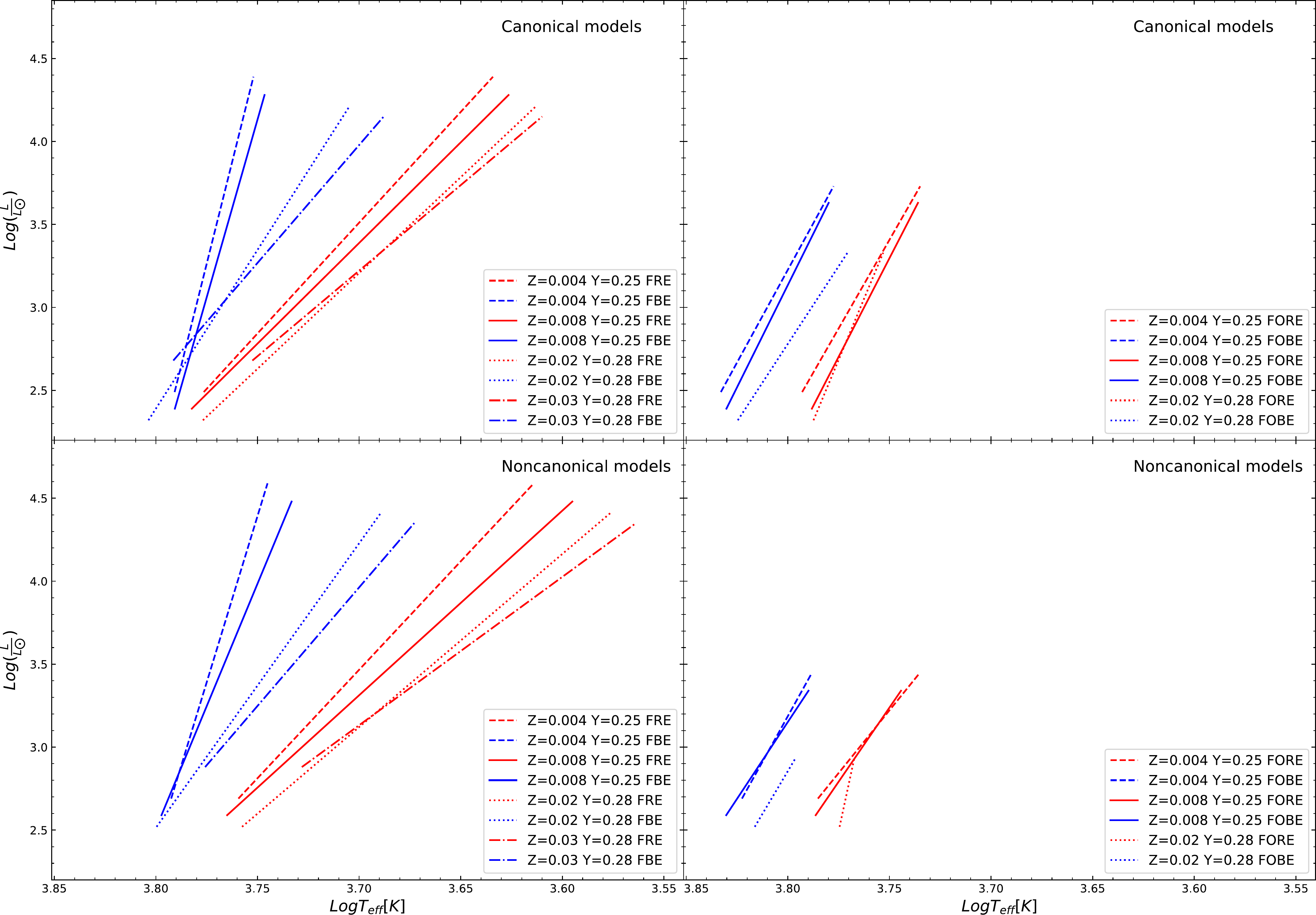}
\caption{{\sl Upper panels}:The F (left panel) and, when available, FO (right panel) instability strip boundaries for $Z$ = $0.004$ (dashed lines), $Z$ = $0.008$ (solid lines), $Z$ = $0.02$ (dotted lines) and $Z$ = $0.03$ (dash-dotted lines), obtained by adopting the canonical ML relation and the standard efficiency for the superadiabatic convection. {\sl Bottom panels}: The same as in the upper panels but for noncanonical models.}
\label{fig:strips_metallicity}
\end{figure*}

\subsection{The pulsational framework}

To investigate the metallicity effect on the pulsational properties of CCs, as well as to pave the way for future theoretical calibrations of the Cepheid-based extragalactic distance scale, a theoretical analysis of the CC pulsation scenario was extended to chemical compositions different from the solar one already analyzed in \citetalias{Desomma2020a}. 
In particular, we considered the following chemical compositions: $Z$=$0.004$, $Y$= $0.25$; $Z$=$0.008$, $Y$= $0.25$; and $Z$=$0.03$, $Y$= $0.28$, to be representative of CCs observed in the Small Magellanic Cloud, the Large Magellanic Cloud, and M31, respectively.\\
For each chemical composition, a wide range of masses (3<\msun<11) and effective temperatures ($3600<T_{eff}$[K]$<7200$), as well as two assumptions of the ML relation were considered. Models with a canonical luminosity level, i.e. models computed by neglecting core convective overshooting, rotation, and mass loss, were labeled `case A`, while those with a noncanonical luminosity level, obtained by increasing the canonical luminosity level by $\Delta\log(L$/$L_\odot)$=$~0.2 \; dex$, were labeled `case B`. An analysis of the impact of a change in the super-adiabatic convection efficiency was also performed for consistency with the analysis performed at solar chemical abundance \citepalias{Desomma2020a}.\\
This notwithstanding, we note that for the present analysis, a standard convective efficiency value (i.e. $\alpha_{ml}$ = 1.5) was accounted for because both the PA and PAC relations were found to be barely affected by the choice of the mixing length parameter \citepalias[see][for more details]{Desomma2020b}.

The adopted input parameters are listed in Table \ref{param_puls_models}. For each selected model and pulsation mode, the system of nonlinear dynamical and convective equations was integrated until a stable limit cycle of pulsation was achieved.\\

The relation of the effective temperature of each IS boundary as a function of the luminosity, obtained by performing a linear fit, for both the F and FO pulsators, is provided in Tables \ref{tl_lin_smc}, to \ref{tl_lin_m31}, for $Z$=$0.004$, $Z$=$0.008$ and $Z$=$0.03$, respectively.

The linear relations for the F and FO-mode (when it was stable) boundaries obtained for the canonical ML relation and the standard convective efficiency are plotted in Fig. \ref{fig:strips_metallicity}. For the sake of comparison, the same figure also shows the IS boundaries for the solar chemical composition ($Z$=$0.02$, $Y$=$0.28$). These plots show that, in agreement with previous studies \citep[][]{BCCM1999,BCC1997}, a change in the chemical composition does significantly affect the topology of the instability strip. As the metallicity increases from $Z$=$0.004$ to $Z$=$0.03$, at a fixed mixing length parameter and ML relation, the IS gets redder. This occurrence is due to both the decreased hydrogen abundance that makes pulsation less efficient at the blue edge of the IS, delaying the onset of pulsation to lower effective temperatures, and the increased contribution of iron bump opacity to pulsation at the IS red boundary, thereby delaying the quenching of pulsation due to convection.\\

Following the same approach adopted for the solar chemical composition case, a linear regression of the period as a function of the luminosity, the mass, and the effective temperature (PMLT relation) was carried out. The coefficients of the relations for F and FO pulsators, for each chemical composition, are reported in Table \ref{pmlt_allZ_f_fo}. A quick inspection of their values clearly shows that the metallicity has a larger effect on the zero-point and on the term related to the effective temperature than on the other coefficients. In agreement with previous investigations \citep[see e.g.][]{BCM2000b}, longer periods are expected for higher metal abundance models, when all the other parameters are fixed.

\onecolumn
\begin{longtable}{ccccccc}
\caption{\normalsize{\label{param_puls_models}}The intrinsic stellar parameters for the cases $Z$=$0.004$, $Z$=$0.008$, $Z$=$0.02$ and $Z$=$0.03$ of the F and FO-mode pulsation models adopted for computing the PA and PAC relations. The various columns list the metallicity, helium content, mass, luminosity level, the ML label, and the pulsation mode.}\\
\hline\hline
Z & Y & \msun & \lsun & ML & mode \\
\hline
\endfirsthead
\caption{continued.}\\
\hline\hline
Z & Y & \msun & \lsun & ML & mode \\
\hline
\endhead
\hline
0.004 & 0.25 & 4.0 & 2.91 & A & F/FO \\
0.004 & 0.25 & 4.0 & 3.11 & B & F/FO \\
0.004 & 0.25 & 5.0 & 3.24 & A & F/FO \\
0.004 & 0.25 & 5.0 & 3.44 & B & F/FO \\
0.004 & 0.25 & 6.0 & 3.50 & A & F/FO \\
0.004 & 0.25 & 6.0 & 3.70 & B & F \\
0.004 & 0.25 & 7.0 & 3.73 & A & F/FO \\
0.004 & 0.25 & 7.0 & 3.93 & B & F \\
0.004 & 0.25 & 8.0 & 3.92 & A & F \\
0.004 & 0.25 & 8.0 & 4.12 & B & F \\
0.004 & 0.25 & 9.0 & 4.09 & A & F \\
0.004 & 0.25 & 9.0 & 4.29 & B & F \\
0.004 & 0.25 & 10.0 & 4.25 & A & F \\
0.004 & 0.25 & 10.0 & 4.45 & B & F \\
0.004 & 0.25 & 11.0 & 4.39 & A & F \\
0.004 & 0.25 & 11.0 & 4.59 & B & F \\
0.008 & 0.25 & 4.0 & 2.81 & A & F/FO \\
0.008 & 0.25 & 4.0 & 3.01 & B & F/FO \\
0.008 & 0.25 & 5.0 & 3.14 & A & F/FO \\
0.008 & 0.25 & 5.0 & 3.34 & B & F/FO \\
0.008 & 0.25 & 6.0 & 3.40 & A & F/FO \\
0.008 & 0.25 & 6.0 & 3.60 & B & F \\
0.008 & 0.25 & 7.0 & 3.63 & A & F/FO \\
0.008 & 0.25 & 7.0 & 3.83 & B & F \\
0.008 & 0.25 & 8.0 & 3.82 & A & F \\
0.008 & 0.25 & 8.0 & 4.02 & B & F \\
0.008 & 0.25 & 9.0 & 3.99 & A & F \\
0.008 & 0.25 & 9.0 & 4.19 & B & F \\
0.008 & 0.25 & 10.0 & 4.14 & A & F \\
0.008 & 0.25 & 10.0 & 4.34 & B & F \\
0.008 & 0.25 & 11.0 & 4.28 & A & F \\
0.008 & 0.25 & 11.0 & 4.48 & B & F \\
0.02 & 0.28 & 4.0 & 2.74 & A & F/FO \\
0.02 & 0.28 & 4.0 & 2.94 & B & F/FO \\
0.02 & 0.28 & 5.0 & 3.07 & A & F/FO \\
0.02 & 0.28 & 5.0 & 3.27 & B & F \\
0.02 & 0.28 & 6.0 & 3.33 & A & F/FO \\
0.02 & 0.28 & 6.0 & 3.53 & B & F \\
0.02 & 0.28 & 7.0 & 3.56 & A & F \\
0.02 & 0.28 & 7.0 & 3.76 & B & F \\
0.02 & 0.28 & 8.0 & 3.75 & A & F \\
0.02 & 0.28 & 8.0 & 3.95 & B & F \\
0.02 & 0.28 & 9.0 & 3.92 & A & F \\
0.02 & 0.28 & 9.0 & 4.12 & B & F \\
0.02 & 0.28 & 10.0 & 4.08 & A & F \\
0.02 & 0.28 & 10.0 & 4.28 & B & F \\
0.02 & 0.28 & 11.0 & 4.21 & A & F \\
0.02 & 0.28 & 11.0 & 4.41 & B & F \\
0.03 & 0.28 & 4.0 & 2.68 & A & F/FO \\
0.03 & 0.28 & 4.0 & 2.88 & B & F \\
0.03 & 0.28 & 5.0 & 3.01 & A & F \\
0.03 & 0.28 & 5.0 & 3.21 & B & F \\
0.03 & 0.28 & 6.0 & 3.27 & A & F \\
0.03 & 0.28 & 6.0 & 3.47 & B & F \\
0.03 & 0.28 & 7.0 & 3.50 & A & F \\
0.03 & 0.28 & 7.0 & 3.70 & B & F \\
0.03 & 0.28 & 8.0 & 3.69 & A & F \\
0.03 & 0.28 & 8.0 & 3.89 & B & F \\
0.03 & 0.28 & 9.0 & 3.86 & A & F \\
0.03 & 0.28 & 9.0 & 4.06 & B & F \\
0.03 & 0.28 & 10.0 & 4.02 & A & F \\
0.03 & 0.28 & 10.0 & 4.22 & B & F \\
0.03 & 0.28 & 11.0 & 4.15 & A & F \\
0.03 & 0.28 & 11.0 & 4.35 & B & F \\
\hline\hline
\end{longtable}
\twocolumn

\begin{table*}
\caption{\label{tl_lin_smc} The coefficients of the relation $\log T_{eff}$ = a +b \lsun, for both the F and FO-mode IS boundaries, for $Z$ = $0.004$ and $Y$ = $0.25$, $\alpha_{ml}$ = 1.5 and various assumptions of the ML relation.  The last two columns represent the root-mean-square deviation ($\sigma$) and the R-squared ($R^2$) coefficients. }
\centering
\begin{tabular}{cccccccccc}
\hline\hline
ML&a&b&$\sigma_{a}$&$\sigma_{b}$&$\sigma$&$R^2$\\
\hline
FBE\\
\hline
A&3.825&-0.016&0.012&0.003&0.005&0.7696\\
B&3.849&-0.022&0.011&0.003&0.005&0.8879\\
\hline
FRE\\
\hline
A&3.958&-0.074&0.014&0.004&0.006&0.9814\\
B&3.937&-0.069&0.033&0.009&0.014&0.9031\\
\hline
FOBE\\
\hline
A&3.941&-0.044&0.012&0.004&0.003&0.9796\\
B&3.945&-0.045&0.005&0.002&0.001&0.9988\\
\hline
FORE\\
\hline
A&3.844&-0.026&0.043&0.013&0.010&0.5588\\
B&3.888&-0.042&0.099&0.032&0.010&0.6320\\
\hline\hline
\end{tabular}
\end{table*}

\begin{table*}
\caption{\label{tl_lin_lmc} The same as in Table \ref{tl_lin_smc} but for $Z$ = $0.008$ and $Y$ = $0.25$}
\centering
\begin{tabular}{ccccccccc}
\hline\hline
ML&a&b&$\sigma_{a}$&$\sigma_{b}$&$\sigma$&$R^2$\\
\hline
FBE\\
\hline
A&3.833&-0.020&0.012&0.003&0.005&0.8362\\
B&3.865&-0.029&0.017&0.005&0.007&0.8444\\
\hline
FRE\\
\hline
A&3.975&-0.081&0.014&0.004&0.006&0.9839\\
B&3.987&-0.087&0.022&0.006&0.009&0.9692\\
\hline
FOBE\\
\hline
A&3.916&-0.037&0.021&0.007&0.005&0.9073\\
B&3.969&-0.054&0.013&0.004&0.001&0.9932\\
\hline
FORE\\
\hline
A&3.864&-0.034&0.029&0.009&0.007&0.8122\\
B&3.879&-0.039&0.078&0.026&0.008&0.6879\\
\hline\hline
\end{tabular}
\end{table*}

\begin{table*}
\caption{\label{tl_lin_m31} The same as in Table \ref{tl_lin_smc} but for $Z$ = $0.03$ and $Y$ = $0.28$.}
\centering
\begin{tabular}{cccccccc}
\hline\hline
ML&a&b&$\sigma_{a}$&$\sigma_{b}$&$\sigma$&$R^2$\\
\hline
FBE\\
\hline
A&3.979&-0.070&0.007&0.002&0.002&0.9953\\
B&3.976&-0.070&0.009&0.002&0.003&0.9937\\
\hline
FRE\\
\hline
A&4.009&-0.096&0.015&0.004&0.005&0.9892\\
B&4.048&-0.111&0.013&0.004&0.004&0.9940\\
\hline\hline
\end{tabular}
\end{table*}
 
\begin{table*}
\caption{\label{pmlt_allZ_f_fo} The coefficients of the PMLT relations $\log P$ = a +b$\log T_{eff}$ + c $\log$ (\msun) + d \lsun for both F and FO pulsators as a function of the assumed $\alpha_{ml}$ parameter for $Z$=$0.004$ and $Y$ = $0.25$, $Z$ = $0.008$ and $Y$ = $0.25$, and $Z$ = $0.03$ and $Y$ = $0.28$. For comparison, the relations for $Z$ = $0.02$ and $Y$ = $0.28$ taken by \citetalias{Desomma2020b} are also reported.}
\centering
\begin{tabular}{cccccccccc}
\hline\hline
a&b&c&d&$\sigma_{a}$&$\sigma_{b}$&$\sigma_{c}$&$\sigma_{d}$&$\sigma$&$R^2$\\
\hline
&&&&Z=0.004 & Y= 0.25\\
\hline
F\\
10.711&-3.315&-0.776&0.918&0.109&0.028&0.017&0.005&0.013&0.9990\\
\hline
FO\\
12.042&-3.636&-0.574&0.799&1.271&0.325&0.112&0.037&0.032&0.9867\\
\hline\hline
&&&&Z=0.008 & Y= 0.25\\
\hline
F\\
10.482&-3.254&-0.773&0.920&0.103&0.026&0.017&0.005&0.013&0.9991\\
\hline
FO\\
10.880&-3.337&-0.622&0.816&0.122&0.031&0.009&0.003&0.003&0.9999\\
\hline\hline
&&&&Z=0.02 & Y= 0.28\\
\hline
F\\
10.268&-3.192&-0.758&0.919&0.001&0.025&0.015&0.005&0.011&0.9995\\
\hline
FO\\
10.595&-3.253&-0.621&0.804&0.002&0.067&0.014&0.005&0.003&0.9996\\
\hline\hline
&&&&Z=0.03 & Y= 0.28\\
\hline
F\\
10.414&-3.227&-0.765&0.918&0.119&0.029&0.023&0.008&0.007&0.9998\\
\hline\hline
\end{tabular}
\end{table*}

\section{The dependence of the period-age and period-age-color relations on the metallicity}

By combining our pulsation model results for the aforementioned chemical compositions with the corresponding stellar evolution model predictions, and adopting the derived PMLT relation, the periods for each combination of mass, luminosity, and effective temperature, along the portions of the evolutionary tracks within the predicted instability strip boundaries (see Fig. \ref{fig:strips_tracks_z}), were derived.\\
A linear regression of the periods estimated from the PMLT relation, the age and mean colors - in the various selected photometric systems - of the corresponding points along the evolutionary tracks, allowed us to derive accurate PA and multi-filter PAC relations, including the first theoretical PAC in the Gaia bands for the new chemical compositions (see Table \ref{param_puls_models}). The coefficients of the PA relation are reported in Table \ref{pa_f_fo_allz}. 

\begin{figure*}
\centering
\includegraphics[width=\textwidth]{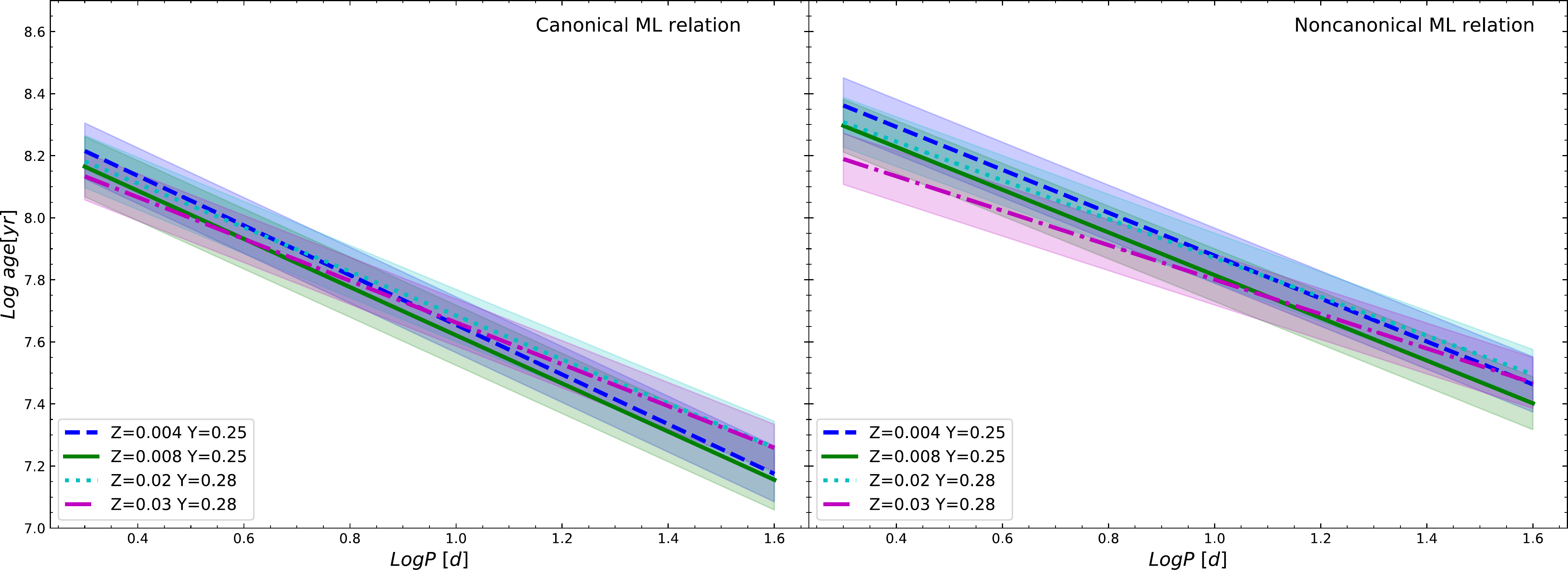}
\caption{Fundamental PA relations for $Z$ = $0.004$ (dashed lines), $Z$ = $0.008$ (solid lines), $Z$ = $0.02$ (dotted lines) and $Z$ = $0.03$ (dash-dotted lines), obtained by adopting the canonical ML relation (left panel) and noncanonical ML relation (right panel) and the standard efficiency for the superadiabatic convection. The colored areas represent the $1\sigma$ errors on these relations.}
\label{fig:pa_f_metallicity}
\end{figure*}

Fig. \ref{fig:pa_f_metallicity} shows the canonical (left panel) and noncanonical (right panel) PA relations obtained for the standard convective efficiency for $Z$ = $0.004$ (dashed blue line), $Z$ = $0.008$ (solid green line), $Z$ = $0.02$ (dotted cyan line) and $Z$ = $0.03$ (dash-dotted magenta line). The colored areas represent the $1\sigma$ errors on these relationships.

For any given initial mass, noncanonical models during the blue loop phase are older and brighter with respect to the canonical ones as a consequence of the larger convective core during the central H-burning stage, and of the larger He core during the central He-burning phase. As a result, PA relations - regardless of the metallicity - based on the noncanonical scenario, predict older ages than the corresponding canonical PA relation as also shown at solar metallicity by \citetalias{Desomma2020b}, for any given period.

Regardless of the selected evolutionary scenario, i.e. either the canonical or the noncanonical one, as the metal abundance increases, the PA relation gets flatter. This occurrence becomes more evident when moving from $Z$=$0.008$ to $Z$=$0.02$ and then to $Z$=$0.03$.

\begin{figure}
\centering
\includegraphics[width=1.1\columnwidth]{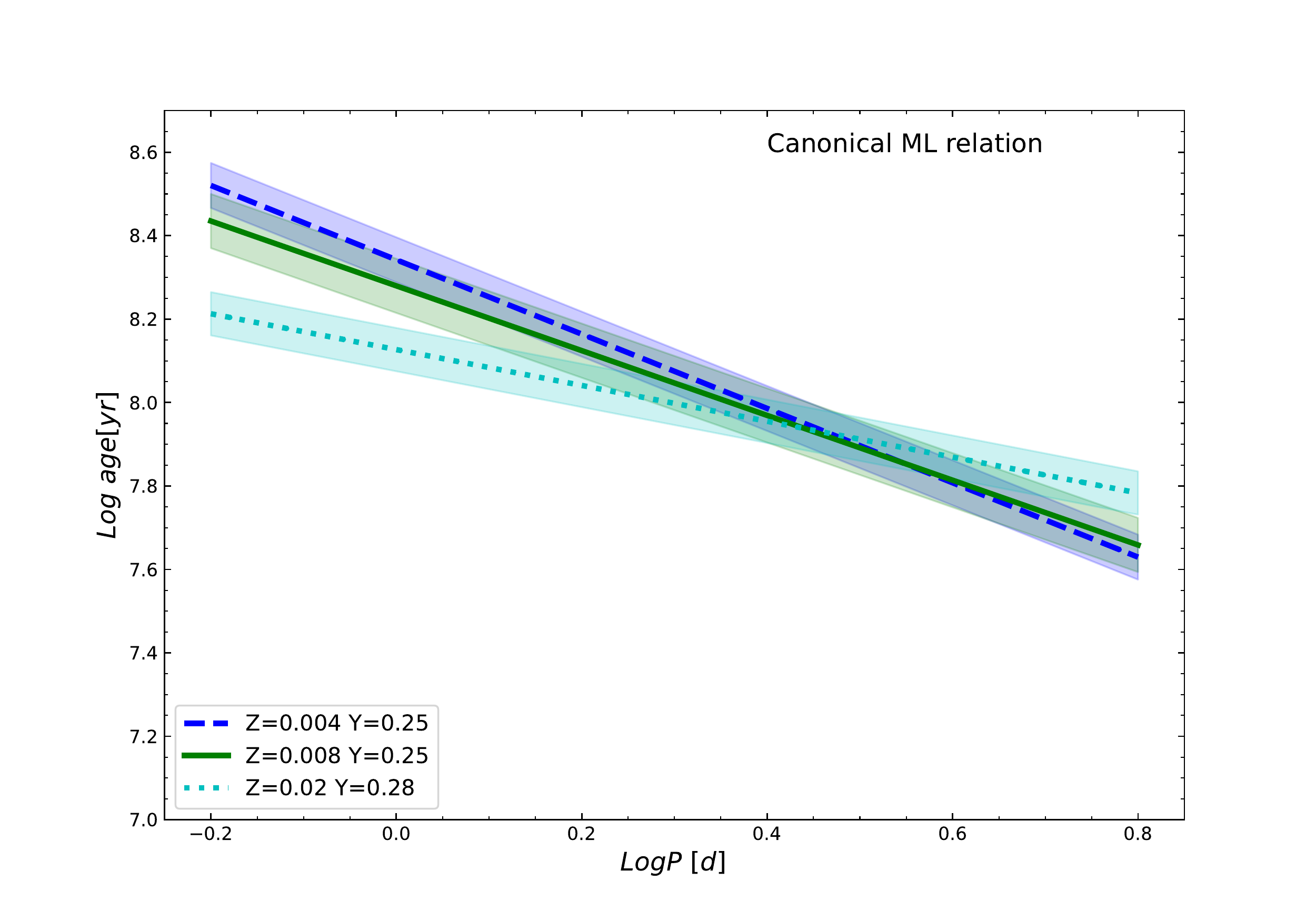}
\caption{First overtone PA relations for $Z$ = $0.004$ (dashed lines), $Z$ = $0.008$ (solid lines), $Z$ = $0.02$ (dotted lines) and $Z$ = $0.03$, obtained by adopting the canonical ML relation and the standard efficiency for the superadiabatic convection.}
\label{fig:pa_fo_metallicity}
\end{figure}

Fig. \ref{fig:pa_fo_metallicity} shows the same comparison as Fig. \ref{fig:pa_f_metallicity} but for FO-mode models. In this case, due to the reduced number of available models, only the canonical cases for $Z$=$0.004$, $Z$=$0.008$ and $Z$=$0.02$ are represented. As FO-mode pulsators have masses and periods smaller than F pulsators, their ages are systematically older than the F ones for each chemical composition. Again, we note a non-negligible metallicity effect on both the slope and the zero point that becomes important, particularly when moving from $Z$=$0.004$  to $Z$=$0.02$. This difference is related to the decreasing number of FO-mode pulsators as the metallicity increases. 

Similar to the PL relation, the PA is affected by the intrinsic dispersion, due to the finite width of the instability strip. This implies that the use of PA relations provides individual CC ages that are affected by systematic errors related to the actual position of the pulsator inside the IS.
In order to take into account this occurrence, as done in \citetalias{Desomma2020b}, we derived the PAC relation for various photometric systems, such as the Gaia EDR3, Johnson-Cousin and JWST NIRCam photometric systems.

The coefficients of the F and FO-mode PAC relation for the Gaia $(G_{BP}-G_{RP})$ color, as well as  $(V-I)$, $(V-K)$, $(J-K)$ and $(F115W-F150W)$, $(F150W-F200W)$ and $(F115W-F200W)$ colors are listed in Tables~\ref{pac_f_allz} and ~\ref{pac_fo_allz}, respectively.\footnote{To validate the inclusion of the color term to the Period-Age relation, an F-test analysis was performed \citep[][and references therein]{Press2003}. Assuming a standard significance level of $\alpha$=$0.05$, and comparing it with the p-values obtained from the F-test, lead us to assert that for the 77 Period-Age-Color relations listed in Tables~\ref{pac_f_allz} and ~\ref{pac_fo_allz} the color term is favoured and therefore, the Period-Age-Color relation is preferred to the Period-Age relation.}

The availability of a theoretical scenario spanning a wide metallicity range allowed us to obtain the Period-Age-Metallicity (PAZ) and Period-Age-Color-Metallicity (PACZ) relations. The coefficients of the PAZ relations are reported in Table~\ref{paz_f_fo_allz} while Table \ref{pacz_f_fo_allz} shows the coefficients of the PACZ relation for both F and FO-mode pulsators. These PAZ and PACZ relations are the first metallicity-dependent theoretical calibration of the CC PA and PAC relations.\footnote{Following a referee's suggestion, we verified that present results are barely affected, if any, by the adopted density grid for both pulsation and evolutionary models. This is because the density of the pulsation model grid has been optimized to properly define the location of the instability strip boundaries. From the point of view of the evolutionary models, the density of the grid - both in terms of the number of evolutionary tracks and time steps for each stellar model - is very fine to properly trace any evolutionary features during the core He-burning stage. This notwithstanding, we decided to perform a test by deriving the PA, PAC, PAZ and PACZ relations after halving the number of the adopted evolutionary models. We found that this assumption increases the errors on the slope and intercept as well as the standard deviation of the relations by about 35\%;  but the average values of the slope and intercept in each relation remain consistent within the errors.}

\begin{table*}
\caption{\label{pa_f_fo_allz}The coefficients of the F
and FO-mode PA relations in the form $\log t$ = $a$ + $b \log P$, for $Z$=$0.004$, $Y$= $0.25$; $Z$=$0.008$, $Y$= $0.25$ and $Z$=$0.03$, $Y$= $0.28$ derived by assuming linear IS boundaries and adopting both {\sl case A} and {\sl B} ML relations. The last two columns represent the root-mean-square deviation ($\sigma$) and the R-squared ($R^2$) coefficients}. For comparison, the relations for $Z$=$0.02$ and $Y$=$0.28$ taken by \citetalias{Desomma2020b} are also reported.
\centering
\begin{tabular}{ccccccc}
\hline\hline
ML&a&b&$\sigma_{a}$&$\sigma_{b}$&$\sigma$&$R^2$\\
\hline
&&Z=0.004 & Y= 0.25\\
\hline
Fundamental mode\\
\hline
A&8.455&-0.800&0.013&0.013&0.090&0.849\\
B&8.570&-0.692&0.019&0.014&0.089&0.767\\
\hline
First overtone mode\\
\hline
A&8.342&-0.891&0.024&0.063&0.054&0.585\\
\hline\hline
&&Z=0.008 & Y= 0.25\\
\hline
Fundamental mode\\
\hline
A&8.398&-0.776&0.016&0.013&0.097&0.681\\
B&8.503&-0.688&0.010&0.007&0.085&0.839\\
\hline
First overtone mode\\
\hline
A&8.280&-0.777&0.023&0.039&0.065&0.770\\
\hline\hline
&&Z=0.02 & Y= 0.28\\
\hline
Fundamental mode\\
\hline
A&8.393&-0.704&0.008&0.009&0.084&0.916\\ 
B&8.480&-0.626&0.010&0.009&0.080&0.866\\
\hline
First overtone mode\\
\hline
A&8.120&-0.396&0.020&0.057&0.052&0.506\\
\hline\hline
&&Z=0.03 & Y= 0.28\\
\hline
Fundamental mode\\
\hline
A&8.336&-0.673&0.009&0.010&0.076&0.887\\
B&8.356&-0.555&0.012&0.009&0.082&0.822\\
\hline\hline
\end{tabular}
\end{table*}

\begin{table*}
\caption{\label{pac_f_allz}The coefficients of the F-mode canonical and noncanonical PAC relations $\log t$ = $a$ + $b \log P$ + $c$ CI), for $Z$=$0.004$, $Y$= $0.25$; $Z$=$0.008$, $Y$= $0.25$; $Z$=$0.02$, $Y$=$0.28$; and $Z$=$0.03$, $Y$= $0.28$. The last two columns represent the root-mean-square deviation ($\sigma$) and the R-squared ($R^2$) coefficients.}
\centering
\begin{tabular}{cccccccccccc}
\hline\hline
Z&Y&color&ML&a&b&c&$\sigma_{a}$&$\sigma_{b}$&$\sigma_{c}$&$\sigma$&$R^2$\\
\hline
Fundamental mode\\
\hline
0.004&0.25&$G_{BP}-G_{RP}$&A&8.794&-0.571&-0.543&0.036&0.026&0.055&0.082&0.868\\
0.004&0.25&V-I&A&8.808&-0.578&-0.618&0.038&0.026&0.063&0.082&0.868\\
0.004&0.25&V-K&A&8.767&-0.576&-0.254&0.033&0.025&0.025&0.081&0.869\\
0.004&0.25&J-K&A&8.715&-0.579&-0.864&0.027&0.024&0.082&0.081&0.870\\
0.004&0.25&F115W-F150W&A&8.769&-0.583&-1.087&0.032&0.024&0.102&0.080&0.871\\
0.004&0.25&F150W-F200W&A&8.616&-0.574&-1.954&0.020&0.025&0.190&0.081&0.869\\
0.004&0.25&F115W-F200W&A&8.716&-0.579&-0.702&0.027&0.024&0.066&0.081&0.870\\
0.004&0.25&$G_{BP}-G_{RP}$&B&8.517&-0.758&0.135&0.024&0.024&0.035&0.068&0.772\\
0.004&0.25&V-I&B&8.080&-0.394&0.047&0.015&0.014&0.027&0.071&0.421\\
0.004&0.25&V-K&B&8.096&-0.379&0.006&0.013&0.013&0.011&0.071&0.421\\
0.004&0.25&J-K&B&8.103&-0.370&-0.006&0.011&0.012&0.034&0.071&0.421\\
0.004&0.25&F115W-F150W&B&8.159&-0.319&-0.227&0.014&0.012&0.045&0.071&0.423\\
0.004&0.25&F150W-F200W&B&8.097&-0.416&0.253&0.008&0.013&0.072&0.071&0.422\\
0.004&0.25&F115W-F200W&B&8.114&-0.353&-0.048&0.011&0.012&0.028&0.071&0.421\\
0.008&0.25&$G_{BP}-G_{RP}$&A&8.772&-0.313&-0.769&0.026&0.030&0.045&0.085&0.733\\
0.008&0.25&V-I&A&8.728&-0.345&-0.794&0.026&0.030&0.051&0.087&0.724\\
0.008&0.25&V-K&A&8.651&-0.332&-0.321&0.022&0.030&0.020&0.086&0.728\\
0.008&0.25&J-K&A&8.574&-0.333&-1.086&0.018&0.029&0.064&0.086&0.731\\
0.008&0.25&F115W-F150W&A&8.691&-0.327&-1.497&0.022&0.027&0.082&0.084&0.738\\
0.008&0.25&F150W-F200W&A&8.407&-0.330&-2.168&0.015&0.031&0.136&0.087&0.726\\
0.008&0.25&F115W-F200W&A&8.578&-0.323&-0.896&0.018&0.029&0.051&0.085&0.734\\
0.008&0.25&$G_{BP}-G_{RP}$&B&8.783&-0.401&-0.533&0.030&0.030&0.054&0.082&0.846\\
0.008&0.25&V-I&B&8.600&-0.581&-0.216&0.030&0.031&0.062&0.085&0.840\\
0.008&0.25&V-K&B&8.620&-0.526&-0.131&0.024&0.030&0.024&0.084&0.841\\
0.008&0.25&J-K&B&8.608&-0.499&-0.527&0.018&0.029&0.078&0.084&0.842\\
0.008&0.25&F115W-F150W&B&8.772&-0.373&-1.207&0.024&0.026&0.098&0.081&0.850\\
0.008&0.25&F150W-F200W&B&8.512&-0.627&-0.337&0.011&0.031&0.165&0.085&0.839\\
0.008&0.25&F115W-F200W&B&8.633&-0.455&-0.525&0.018&0.029&0.063&0.083&0.844\\
0.02&0.28&$G_{BP}-G_{RP}$&A&8.416&-0.711&-0.019&0.037&0.022&0.049&0.080&0.917\\
0.02&0.28&V-I&A&8.338&-0.740&0.090&0.040&0.023&0.062&0.080&0.921\\
0.02&0.28&V-K&A&8.351&-0.738&0.033&0.033&0.023&0.024&0.080&0.921\\
0.02&0.28&J-K&A&8.360&-0.735&0.109&0.027&0.022&0.084&0.080&0.921\\
0.02&0.28&F115W-F150W&A&8.347&-0.736&0.156&0.035&0.022&0.115&0.080&0.921\\
0.02&0.28&F150W-F200W&A&8.377&-0.735&0.201&0.016&0.023&0.162&0.080&0.921\\
0.02&0.28&F115W-F200W&A&8.360&-0.736&0.089&0.027&0.022&0.068&0.080&0.921\\
0.02&0.28&$G_{BP}-G_{RP}$&B&8.439&-0.702&0.102&0.026&0.017&0.034&0.079&0.858\\
0.02&0.28&V-I&B&8.462&-0.692&0.082&0.023&0.017&0.036&0.079&0.857\\
0.02&0.28&V-K&B&8.473&-0.690&0.031&0.019&0.017&0.014&0.079&0.857\\
0.02&0.28&J-K&B&8.482&-0.686&0.099&0.017&0.016&0.049&0.079&0.857\\
0.02&0.28&F115W-F150W&B&8.457&-0.693&0.182&0.022&0.016&0.068&0.079&0.858\\
0.02&0.28&F150W-F200W&B&8.499&-0.685&0.170&0.012&0.017&0.093&0.079&0.857\\
0.02&0.28&F115W-F200W&B&8.478&-0.690&0.091&0.017&0.016&0.039&0.079&0.857\\
0.03&0.28&$G_{BP}-G_{RP}$&A&8.211&-0.752&0.181&0.036&0.024&0.050&0.074&0.890\\
0.03&0.28&V-I&A&8.227&-0.746&0.186&0.033&0.023&0.054&0.074&0.890\\
0.03&0.28&V-K&A&8.251&-0.743&0.071&0.027&0.023&0.021&0.074&0.890\\
0.03&0.28&J-K&A&8.268&-0.739&0.247&0.022&0.022&0.075&0.074&0.889\\
0.03&0.28&F115W-F150W&A&8.238&-0.742&0.359&0.029&0.022&0.102&0.074&0.890\\
0.03&0.28&F150W-F200W&A&8.306&-0.738&0.443&0.013&0.023&0.142&0.074&0.889\\
0.03&0.28&F115W-F200W&A&8.268&-0.741&0.201&0.022&0.022&0.060&0.074&0.890\\
0.03&0.28&$G_{BP}-G_{RP}$&B&8.369&-0.539&-0.027&0.021&0.023&0.035&0.080&0.822\\
0.03&0.28&V-I&B&8.374&-0.526&-0.050&0.017&0.022&0.034&0.080&0.822\\
0.03&0.28&V-K&B&8.368&-0.530&-0.018&0.015&0.022&0.014&0.080&0.822\\
0.03&0.28&J-K&B&8.365&-0.529&-0.068&0.014&0.021&0.050&0.080&0.822\\
0.03&0.28&F115W-F150W&B&8.368&-0.538&-0.067&0.018&0.021&0.073&0.080&0.822\\
0.03&0.28&F150W-F200W&B&8.354&-0.530&-0.118&0.012&0.022&0.095&0.080&0.822\\
0.03&0.28&F115W-F200W&B&8.363&-0.535&-0.044&0.014&0.021&0.041&0.080&0.822\\
\hline\hline
\end{tabular}
\end{table*}

\begin{table*}
\caption{\label{pac_fo_allz} The same as in Table \ref{pac_f_allz} but for FO-mode canonical PAC relations.}
\centering
\begin{tabular}{cccccccccccc}
\hline\hline
Z&Y&color&ML&a&b&c&$\sigma_{a}$&$\sigma_{b}$&$\sigma_{c}$&$\sigma$&$R^2$\\
\hline
First overtone mode\\
\hline
0.004&0.25&$G_{BP}-G_{RP}$&A&8.002&-1.097&0.668&0.107&0.088&0.206&0.047&0.614\\
0.004&0.25&V-I&A&7.995&-1.088&0.741&0.110&0.087&0.230&0.047&0.613\\
0.004&0.25&V-K&A&8.005&-1.094&0.333&0.107&0.088&0.103&0.047&0.614\\
0.004&0.25&J-K&A&8.061&-1.089&1.173&0.090&0.087&0.363&0.047&0.613\\
0.004&0.25&F115W-F150W&A&8.054&-1.071&1.266&0.094&0.084&0.399&0.047&0.612\\
0.004&0.25&F150W-F200W&A&8.034&-1.125&4.024&0.096&0.093&1.211&0.046&0.615\\
0.004&0.25&F115W-F200W&A&8.049&-1.084&0.964&0.094&0.086&0.300&0.047&0.613\\
0.008&0.25&$G_{BP}-G_{RP}$&A&8.044&-0.601&0.180&0.100&0.076&0.179&0.064&0.421\\
0.008&0.25&V-I&A&8.040&-0.599&0.208&0.104&0.075&0.207&0.064&0.421\\
0.008&0.25&V-K&A&8.042&-0.601&0.095&0.100&0.075&0.092&0.064&0.422\\
0.008&0.25&J-K&A&8.058&-0.599&0.336&0.085&0.074&0.326&0.064&0.422\\
0.008&0.25&F115W-F150W&A&8.055&-0.594&0.367&0.090&0.072&0.364&0.064&0.421\\
0.008&0.25&F150W-F200W&A&8.050&-0.611&1.161&0.087&0.079&1.044&0.064&0.422\\
0.008&0.25&F115W-F200W&A&8.054&-0.598&0.279&0.089&0.074&0.270&0.064&0.422\\
0.02&0.28&$G_{BP}-G_{RP}$&A&8.082&-0.452&0.069&0.142&0.097&0.223&0.050&0.332\\
0.02&0.28&V-I&A&8.082&-0.451&0.078&0.144&0.095&0.258&0.050&0.331\\
0.02&0.28&V-K&A&8.083&-0.452&0.036&0.135&0.095&0.115&0.050&0.332\\
0.02&0.28&J-K&A&8.090&-0.451&0.131&0.115&0.094&0.416&0.050&0.332\\
0.02&0.28&F115W-F150W&A&8.090&-0.448&0.139&0.121&0.090&0.471&0.050&0.331\\
0.02&0.28&F150W-F200W&A&8.094&-0.452&0.364&0.108&0.100&1.224&0.050&0.331\\
0.02&0.28&F115W-F200W&A&8.091&-0.449&0.101&0.118&0.093&0.340&0.050&0.331\\
\hline\hline
\end{tabular}
\end{table*}

\begin{table*}
\caption{\label{paz_f_fo_allz}The coefficients of the F
and FO-mode PAZ relations in the form $\log t$ = $a$ + $b \log P$ + c [Fe/H] obtained by using all the computed models for $Z$ = $0.004$, $Z$ = $0.008$, $Z$= $0.02$ and $Z$ = $0.03$. The root-mean-square deviation ($\sigma$) and the R-squared ($R^2$) coefficients are reported in the last two columns.}
\centering
\begin{tabular}{ccccccccc}
\hline\hline
ML&a&b&c&$\sigma_{a}$&$\sigma_{b}$&$\sigma_{c}$&$\sigma$&$R^2$\\
\hline
Fundamental mode\\
\hline
A&8.419&-0.775&-0.015&0.006&0.006&0.007&0.083&0.845\\
B&8.423&-0.642&-0.067&0.006&0.004&0.006&0.081&0.842\\
\hline
First overtone mode\\
\hline
A&8.236&-0.762&-0.079&0.014&0.028&0.015&0.041&0.686\\
\hline\hline
\end{tabular}
\end{table*}

\begin{table*}
\caption{\label{pacz_f_fo_allz}The coefficients of the F
and FO PACZ relations $\log t$ = $a$ + $b \log P$ + $c$ CI + d [Fe/H] ), derived by adopting both {\sl cases A} and {\sl B} ML relations.}
\centering
\begin{tabular}{cccccccccccc}
\hline\hline
color&ML&a&b&c&d&$\sigma_{a}$&$\sigma_{b}$&$\sigma_{c}$&$\sigma_{d}$&$\sigma$&$R^2$\\
\hline
Fundamental mode\\
\hline
$G_{BP}-G_{RP}$&A&8.711&-0.559&-0.446&-0.048&0.018&0.014&0.026&0.026&0.042&0.858\\
V-I&A&8.676&-0.577&-0.460&-0.038&0.017&0.014&0.029&0.007&0.078&0.856\\
V-K&A&8.628&-0.572&-0.186&-0.023&0.014&0.013&0.011&0.007&0.078&0.857\\
J-K&A&8.581&-0.570&-0.650&-0.001&0.011&0.013&0.036&0.007&0.077&0.859\\
F115W-F150W&A&8.651&-0.558&-0.921&-0.012&0.013&0.012&0.046&0.007&0.076&0.862\\
F150W-F200W&A&8.490&-0.589&-1.177&0.030&0.007&0.013&0.075&0.007&0.079&0.856\\
F115W-F200W&A&8.584&-0.566&-0.530&0.006&0.011&0.013&0.029&0.007&0.077&0.859\\
$G_{BP}-G_{RP}$&B&8.560&-0.519&-0.231&-0.014&0.013&0.011&0.020&0.020&0.043&0.847\\
V-I&B&8.360&-0.494&0.077&-0.022&0.011&0.010&0.020&0.005&0.080&0.670\\
V-K&B&8.374&-0.467&0.054&-0.021&0.009&0.010&0.008&0.004&0.080&0.671\\
J-K&B&8.366&-0.457&0.215&-0.029&0.007&0.010&0.026&0.004&0.080&0.671\\
F115W-F150W&B&8.318&-0.419&-0.246&-0.004&0.010&0.009&0.036&0.004&0.076&0.617\\
F150W-F200W&B&8.265&-0.450&-0.147&0.008&0.006&0.010&0.054&0.004&0.076&0.616\\
F115W-F200W&B&8.290&-0.430&-0.113&0.003&0.008&0.010&0.022&0.004&0.076&0.617\\

\hline
First overtone mode\\
\hline
$G_{BP}-G_{RP}$&A&7.958&-0.832&0.410&-0.155&0.073&0.033&0.106&0.106&0.040&0.698\\
V-I&A&7.965&-0.832&0.463&-0.134&0.073&0.033&0.125&0.020&0.040&0.698\\
V-K&A&7.979&-0.837&0.208&-0.123&0.070&0.034&0.056&0.018&0.040&0.698\\
J-K&A&8.023&-0.836&0.734&-0.110&0.059&0.034&0.202&0.016&0.040&0.697\\
F115W-F150W&A&8.022&-0.827&0.798&-0.105&0.061&0.033&0.226&0.016&0.040&0.697\\
F150W-F200W&A&8.003&-0.853&2.462&-0.121&0.061&0.036&0.639&0.018&0.040&0.699\\
F115W-F200W&A&8.016&-0.833&0.605&-0.109&0.061&0.034&0.167&0.016&0.040&0.697\\
\hline\hline
\end{tabular}
\end{table*}

\subsection{Comparison with literature results}

\begin{figure}
\includegraphics[width=\columnwidth]{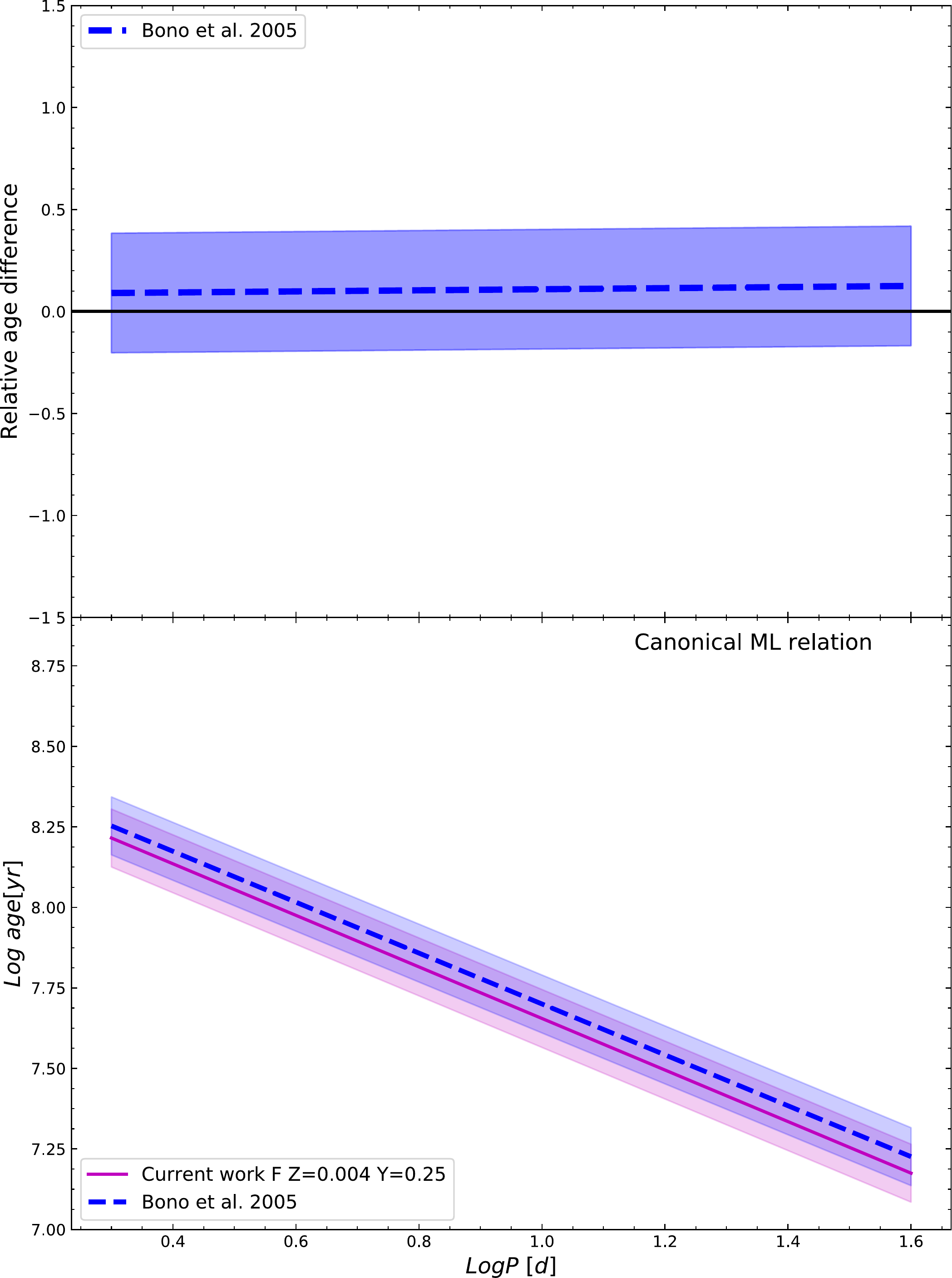}\par
\caption{{\sl Bottom panel}: comparison between F-mode PA
relations derived in the present work for $Z$=$0.004$ (solid magenta line); obtained by varying the adopted ML relation (see labels) with the canonical PA relations derived by \citet{Bono2005}. The colored areas represent the $1\sigma$ errors on these relationships. {\sl Upper panel}: the relative difference, with the expected uncertainty, between the age predictions obtained from the PA relationships compared in the bottom panel.}
\label{fig:pa_smc_F_A_comparison}
\end{figure}

\begin{figure}
\includegraphics[width=\columnwidth]{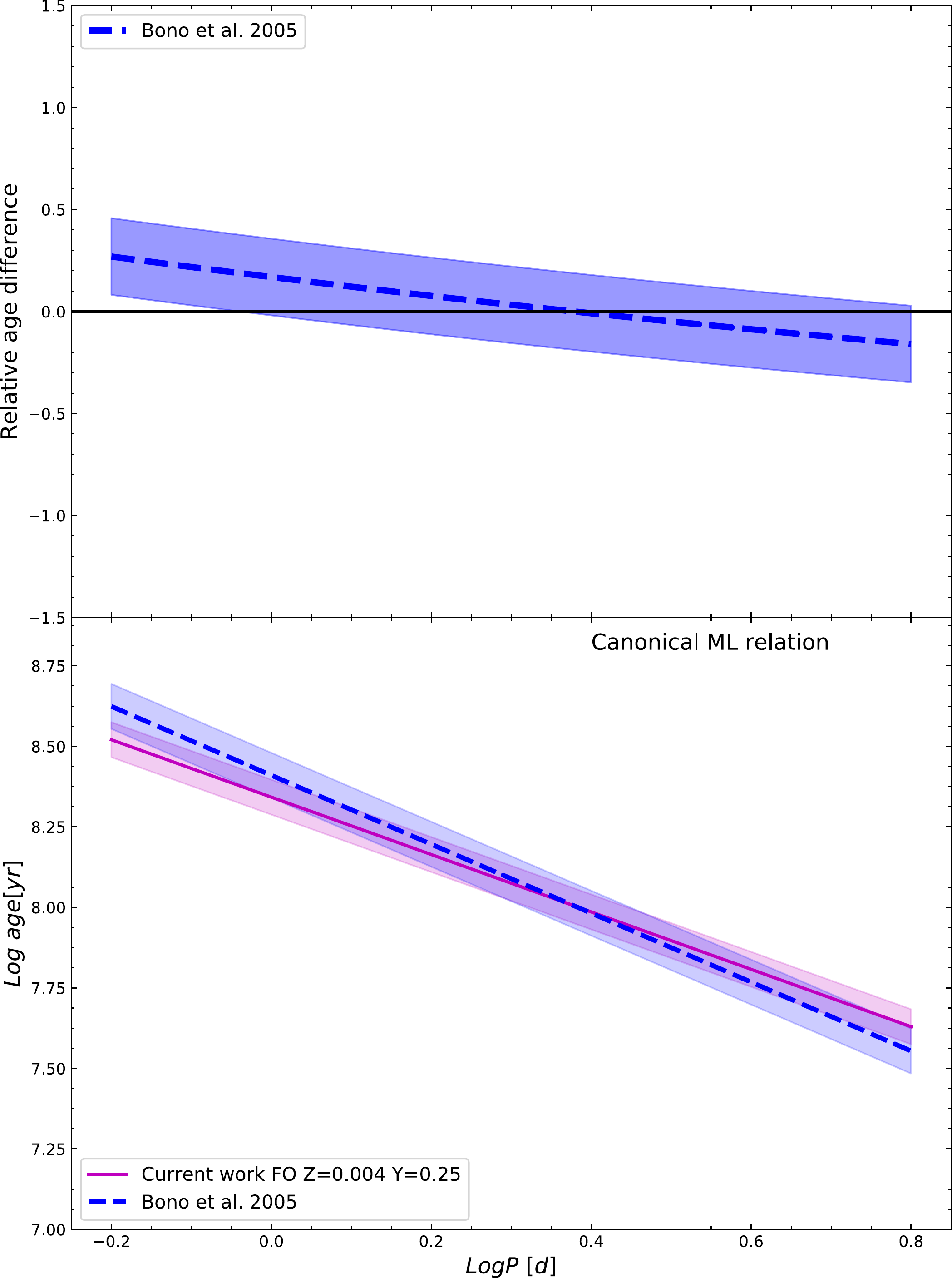}\par
\caption{The same as in Fig. \ref{fig:pa_smc_F_A_comparison} but for FO-mode pulsators.}
\label{fig:pa_smc_FO_A_comparison}
\end{figure}

In this Section we compare the derived PA and PAC relations with those obtained by other authors (see Table 4 in \citetalias{Desomma2020b} for more details). For the chemical composition $Z$ = $0.004$ and $Y$ = $0.25$, Fig. \ref{fig:pa_smc_F_A_comparison} shows the comparison between our canonical PA relation derived for F-mode pulsators (magenta solid line) with the ones derived by \citet{Bono2005} (blue dashed line). Fig. \ref{fig:pa_smc_FO_A_comparison} shows the same comparison but for FO-mode pulsators. 

These figures show that, for F pulsators, within the error, a good agreement does exist, whereas in the case of the FO-mode, our relation is steeper than that provided by \citet{Bono2005}; although the two slopes are still consistent within the errors. This difference can be related to, the small number of FO pulsators with respect to the F-mode models, a slight difference in the assumed ML relation, and the selected mass range because we only selected masses equal to or greater than four solar masses while in \citet{Bono2005} masses less than four solar masses were also included.

\begin{figure*}
\begin{multicols}{2}
\includegraphics[width=\linewidth]{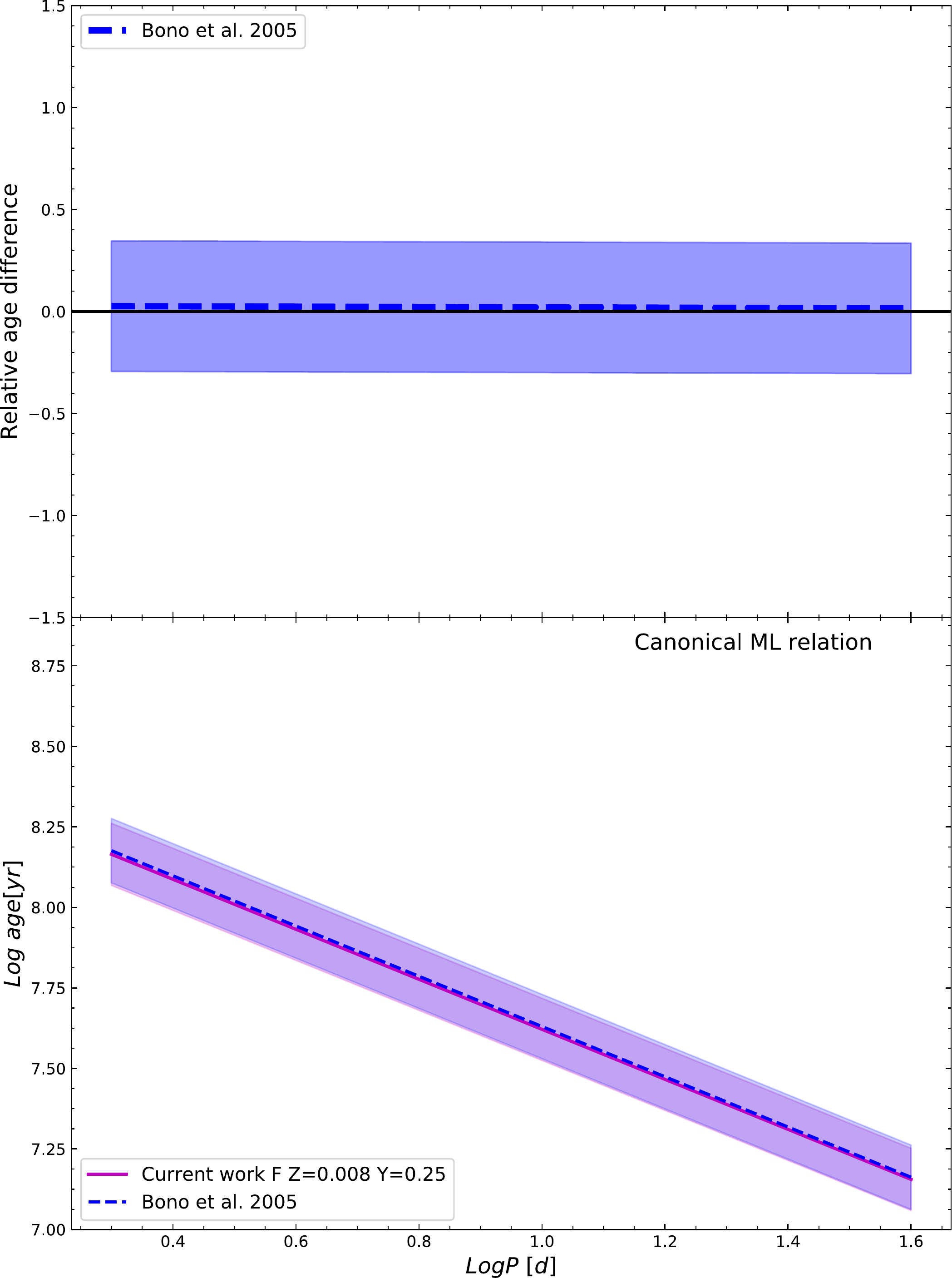}\par 
\includegraphics[width=\linewidth]{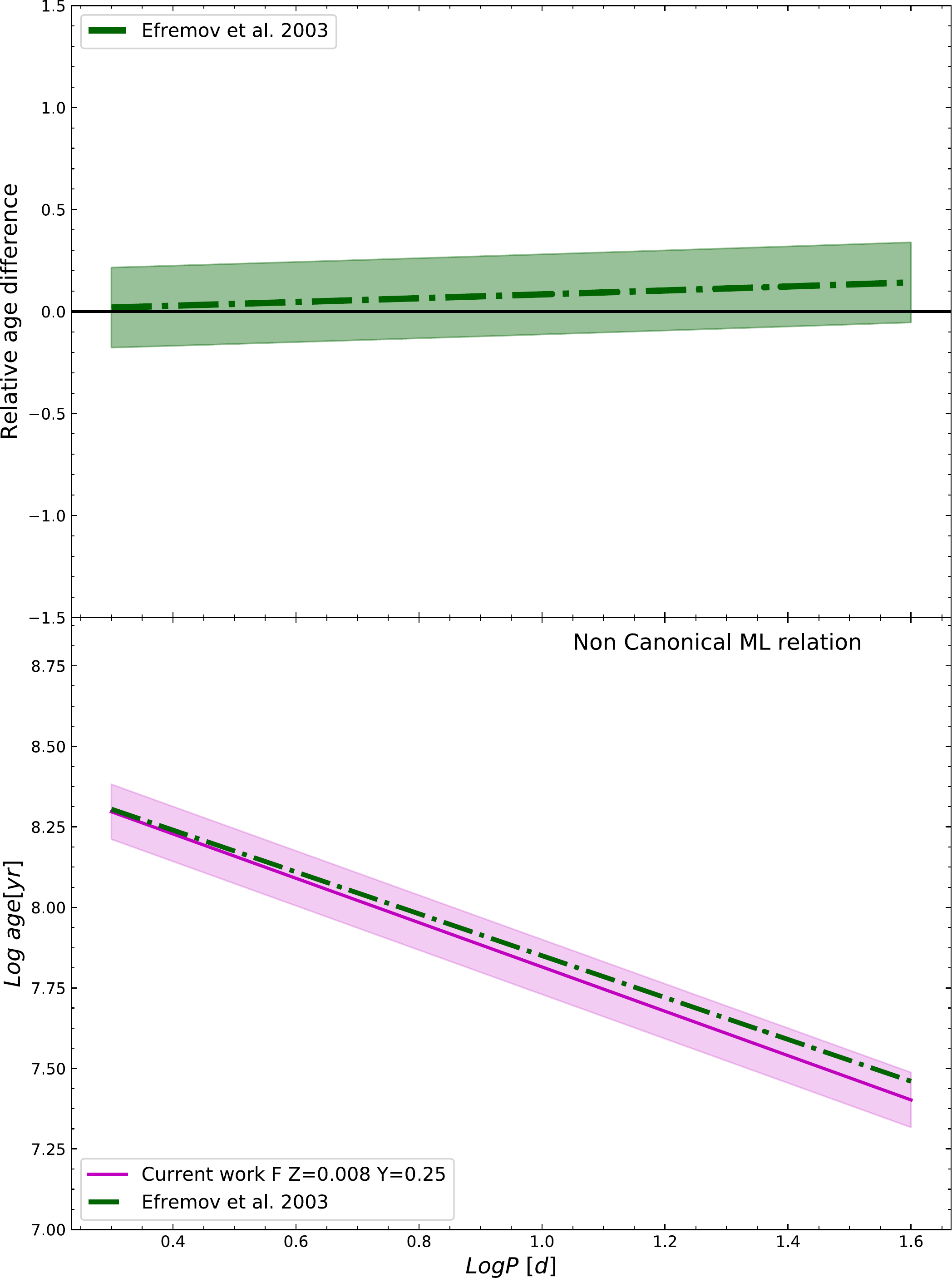}\par
\end{multicols}
\caption{\label{fig:pa_lmc_comparison} {\sl Bottom panels}: comparison between present F-mode PA
relations obtained for $Z$=$0.008$ (solid magenta line) by varying the adopted ML relation (see labels), with similar predictions from the literature: the dashed blue line shows the canonical PA relations by \citet{Bono2005} (left panel) and the dashed green line the noncanonical PA relation by \citet{Efremov2003} (right panel). The $1\sigma$ errors on these relationships as provided by the authors are represented with the colored areas. {\sl Upper panels}: the relative age difference between the age predictions obtained by current PA relations for the canonical case {\sl (left panel)} and the noncanonical one {\sl (right panel)} and those obtained with the PA relations taken from the literature.}
\end{figure*}

In Fig. \ref{fig:pa_lmc_comparison}, the left panel displays the comparison between our canonical PA relations obtained for $Z$=$0.008$ and $Y$=$0.25$ (magenta solid line) and the one derived by \citet{Bono2005} (blue dashed line), while the right panel shows our noncanonical PA relation for the same chemical composition compared with the \citet{Efremov2003} results (green dash-dotted line). Our relation for $Z$=$0.008$ is in excellent accord with the result by \citet{Bono2005}, and slightly steeper than the \citet{Efremov2003} relation, but still consistent within the corresponding errors. 

\begin{figure}
\includegraphics[width=\columnwidth]{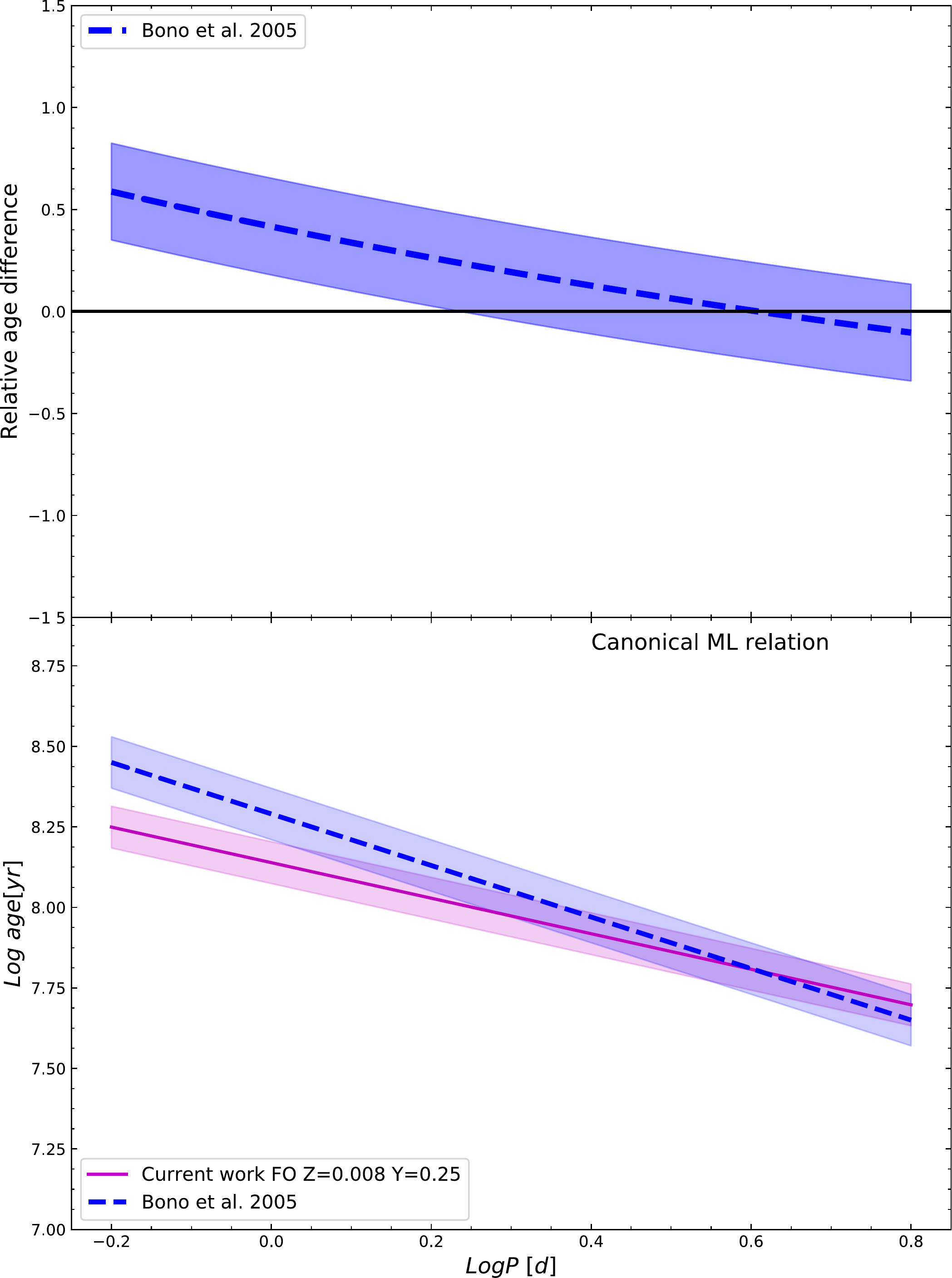}\par
\caption{{\sl Bottom panel}: comparison between present FO-mode PA relation (solid magenta line) obtained for $Z$=$0.008$ and ML case A, and the theoretical FO-mode GCC PA relation obtained by \citet{Bono2005} (dashed blue line). {\sl Upper panel}: the relative difference between the age estimates provided by these PA relations.}
\label{fig:pa_lmc_FO_comparison}
\end{figure}

\begin{figure}
\includegraphics[width=\columnwidth]{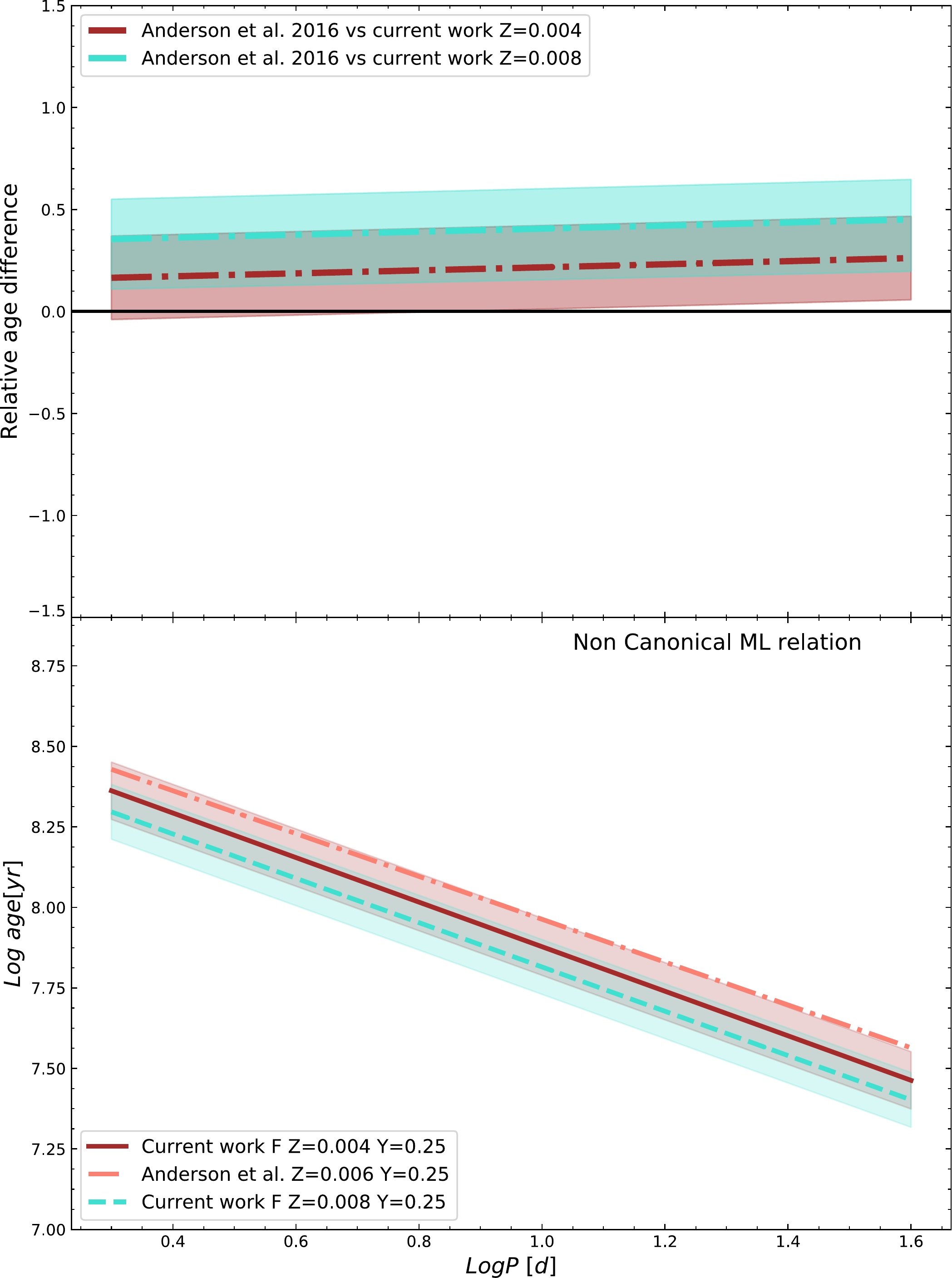}\par
\caption{{\sl Bottom panel}: comparison between F-mode noncanonical PA relations for $Z$=$0.004$ (solid brown line) and $Z$ = $0.008$ (dashed line) derived in the present work and the noncanonical PA relation provided by \citet{Anderson2016} for $Z$ = $0.006$. {\sl Upper panel}: the relative difference between the age estimates provided by these PA relations.}
\label{fig:pa_smc_lmc_And_F_comparison}
\end{figure}

The comparison between our canonical PA relation for FO-mode pulsators with $Z$= $0.008$ and the PA relation by \citet{Bono2005} is shown in Fig. \ref{fig:pa_lmc_FO_comparison}. Here, the difference in the slope is still more significant than at $Z$=$0.004$ because of the smaller number of FO-mode pulsators as the metallicity increases. In Fig. \ref{fig:pa_smc_lmc_And_F_comparison}, we compare our noncanonical PA relation for $Z$=$0.004$ and $Z$=$0.008$ with the relation derived by \citet{Anderson2016} for $Z$=$0.006$. The \citet{Anderson2016} PA relation provides ages that are systematically older than the ones predicted by our relations. This discrepancy is due to the brighter noncanonical ML relation adopted by \citet{Anderson2016} which includes both overshooting and rotation in their models. 

\begin{figure}
\includegraphics[width=\columnwidth]{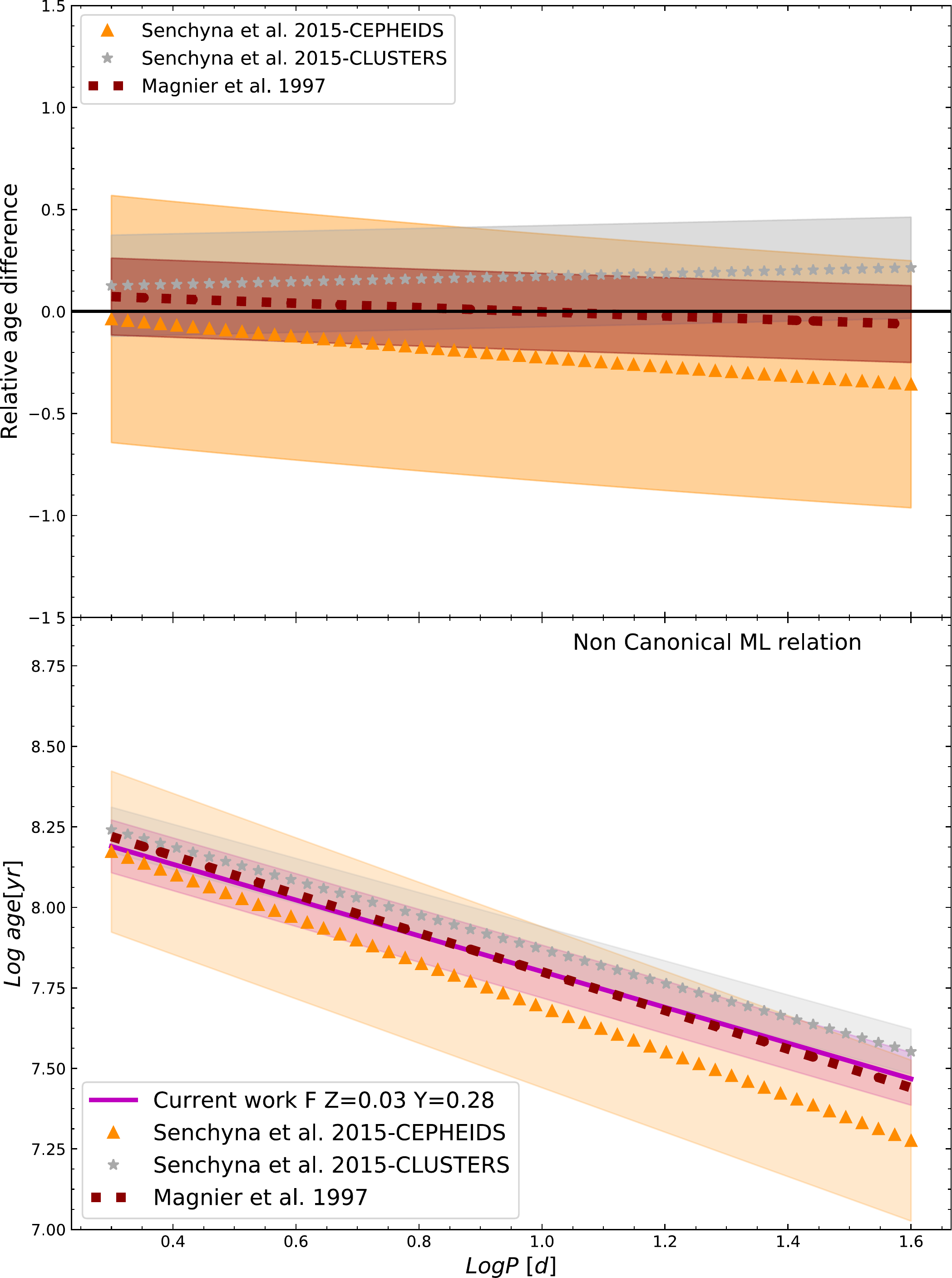}\par
\caption{{\sl Bottom panel}: comparison between present noncanonical F-mode PA relation (solid magenta line) obtained for $Z$=$0.03$ and ML case B, and the PA relation obtained by \citet{Magnier1997} (dotted red line). The relations marked with triangles and stars represent the PA relations for M31 CCs and M31 clusters, respectively; both derived by \citet{Senchyna2015}. {\sl Upper panel}: the relative age difference provided by these PA relations.}
\label{fig:pa_m31_B_F_comp}
\end{figure}

In Fig. \ref{fig:pa_m31_B_F_comp}, our noncanonical PA relation, derived for $Z$=$0.03$ and $Y$ = $0.28$ (solid magenta line) is compared with the one derived by \citet{Magnier1997} (dotted red line) and the ones derived by \citet{Senchyna2015} for M31 CCs (orange triangle marker) and M31 clusters (grey star marker). While the agreement with \citet{Magnier1997} is very good, our relations show fainter/brighter zero points than the \citet{Senchyna2015} results based on cluster Cepheids. In any case, their associated errors are so large that our predictions are still consistent with the results by these authors. The upper panels of Figs. \ref{fig:pa_smc_F_A_comparison}-\ref{fig:pa_m31_B_F_comp} show the relative age difference between our PA relations and the literature ones (see labels).

\section{Individual ages of Gaia Early Data Release 3 Classical Cepheids}
In this Section, we apply the theoretical relations discussed above to a subsample of Gaia Early Data Release 3 (EDR3) Galactic Cepheids with bayesian distances \citep[see][for more details]{Bailer-Jones2021} and complementary metal abundance data \citep{Gaia_Clementini2017, Groenewegen2018, Ripepi2020}, in order to apply both the PAZ, PACZ relations and the PA, PAC, at a fixed solar metallicity, to the Cepheid sample.
The resulting sample consists of 416 F-mode and 56 FO-mode pulsators.

\subsection{Individual age estimates from PAZ and PACZ relations}

\begin{figure*}
\begin{multicols}{2}
    \includegraphics[width=\linewidth]{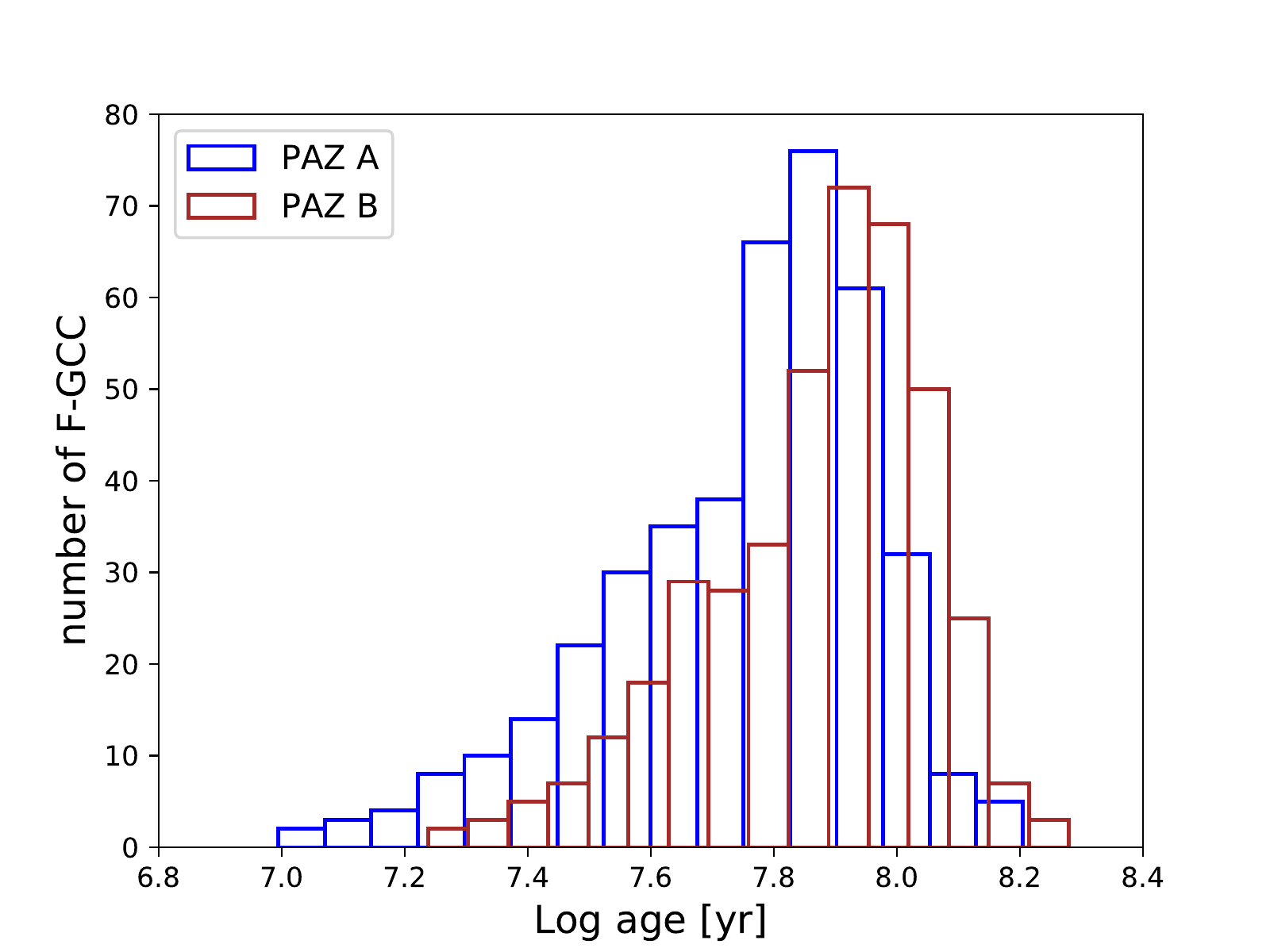}\par 
    \includegraphics[width=\linewidth]{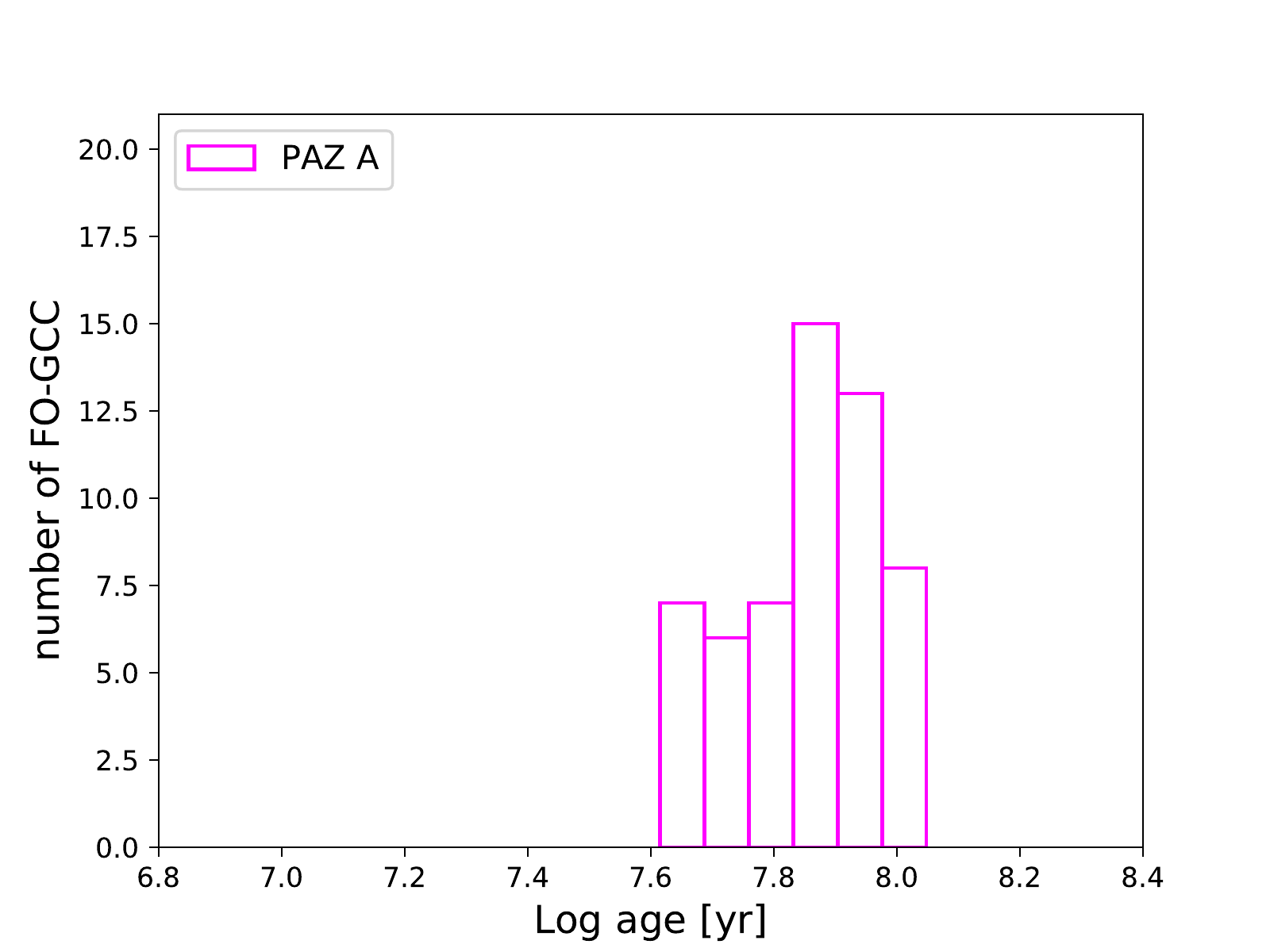}\par 
\end{multicols}
\caption{\label{fig:hist_PAZ} The predicted age distribution, derived by applying the PAZ relation to the selected sample of F {\sl (left panel)} and FO-mode {\sl (right panel)} CCs, for the assumed cases A and B of the ML relation (see labels).}
\end{figure*}

\begin{figure*}
\begin{multicols}{2}
    \includegraphics[width=\linewidth]{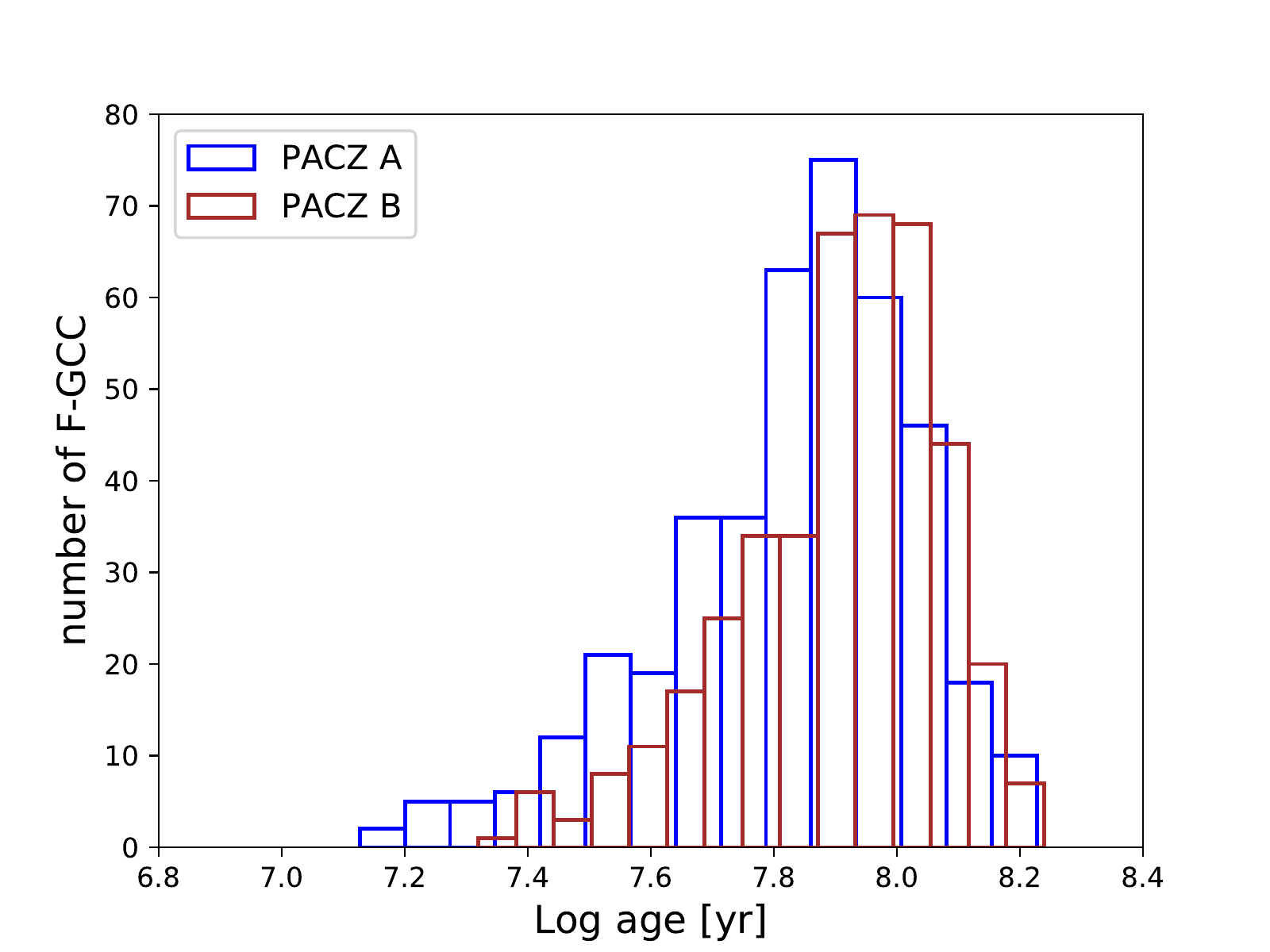}\par 
    \includegraphics[width=\linewidth]{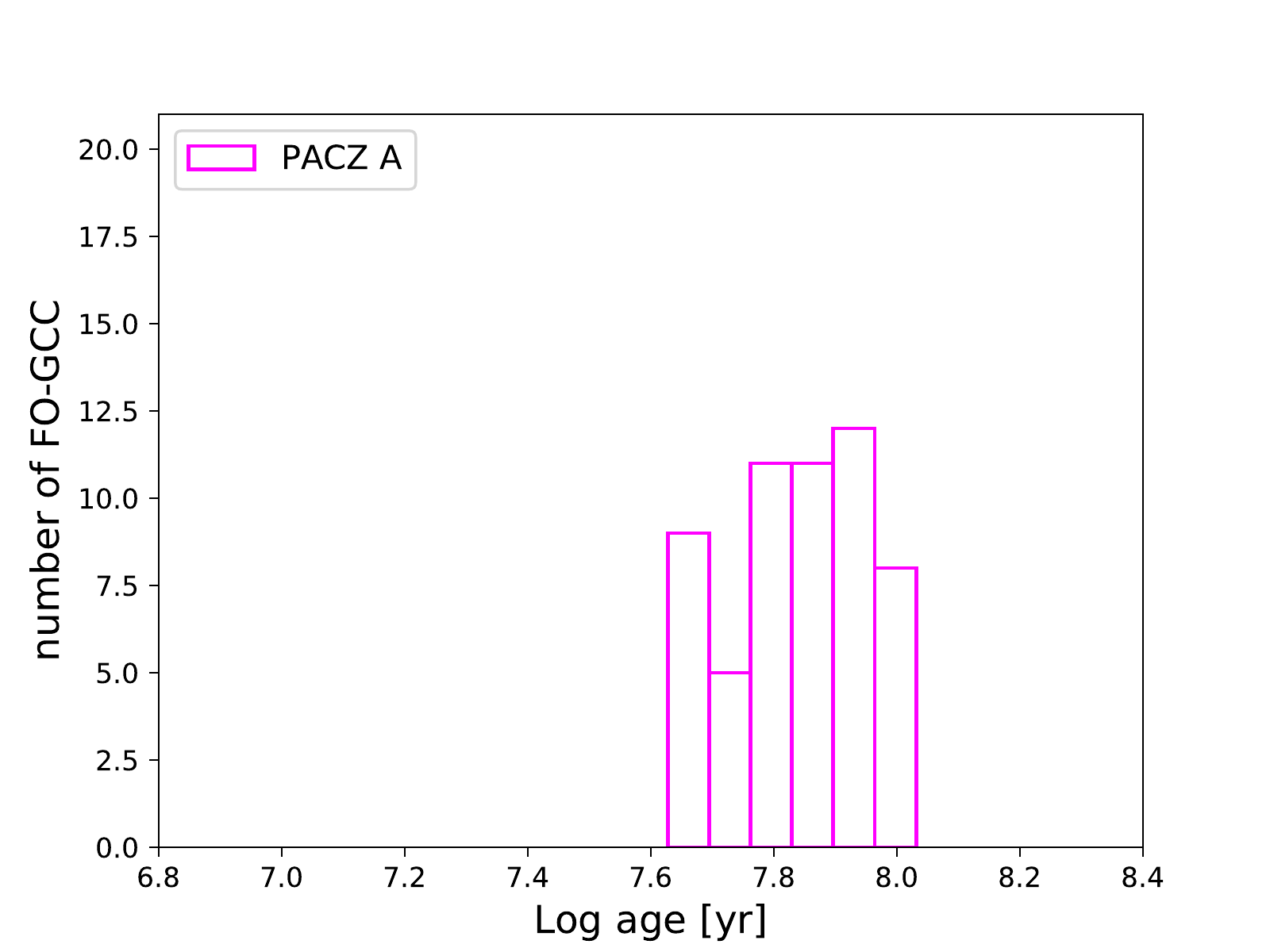}\par 
\end{multicols}
\caption{\label{fig:hist_PACZ} The predicted age distribution, derived by applying the PACZ relation to the selected sample of F {\sl (left panel)} and FO-mode {\sl (right panel)} CCs, for the assumed cases A and B of the ML relation (see labels).}
\end{figure*}

By applying the PAZ and PACZ relations derived above, we estimated the individual ages of all the Cepheids in the selected sample. Figs. \ref{fig:hist_PAZ} and \ref{fig:hist_PACZ} show the distribution of the individual ages computed by using PAZ and PACZ relations, respectively. Inspection of these plots suggests that, in both cases, the age distribution depends on the assumed  ML relation, with the brighter ML relation (case B) providing a shift of the age distribution towards older ages. This trend confirms previous results by \citet{Anderson2016}, \citet{Bono2005} and \citetalias{Desomma2020b}. Furthermore, a comparison between Fig. \ref{fig:hist_PAZ} and \ref{fig:hist_PACZ} confirms that similar results for the main peak of the age distribution are inferred when using the PAZ or the PACZ relations. 

\begin{figure*}
\begin{multicols}{2}
    \includegraphics[width=\linewidth]{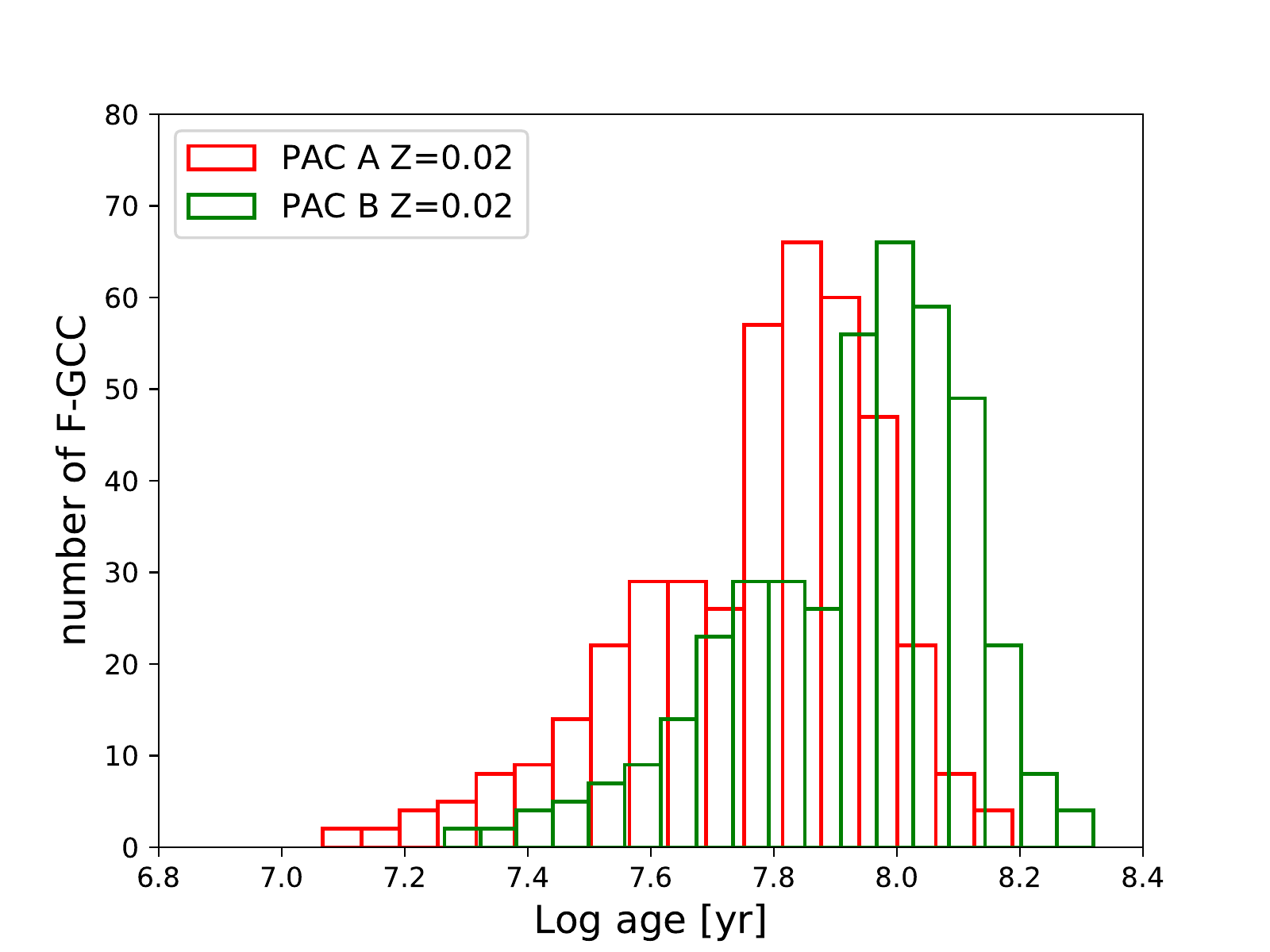}\par 
    \includegraphics[width=\linewidth]{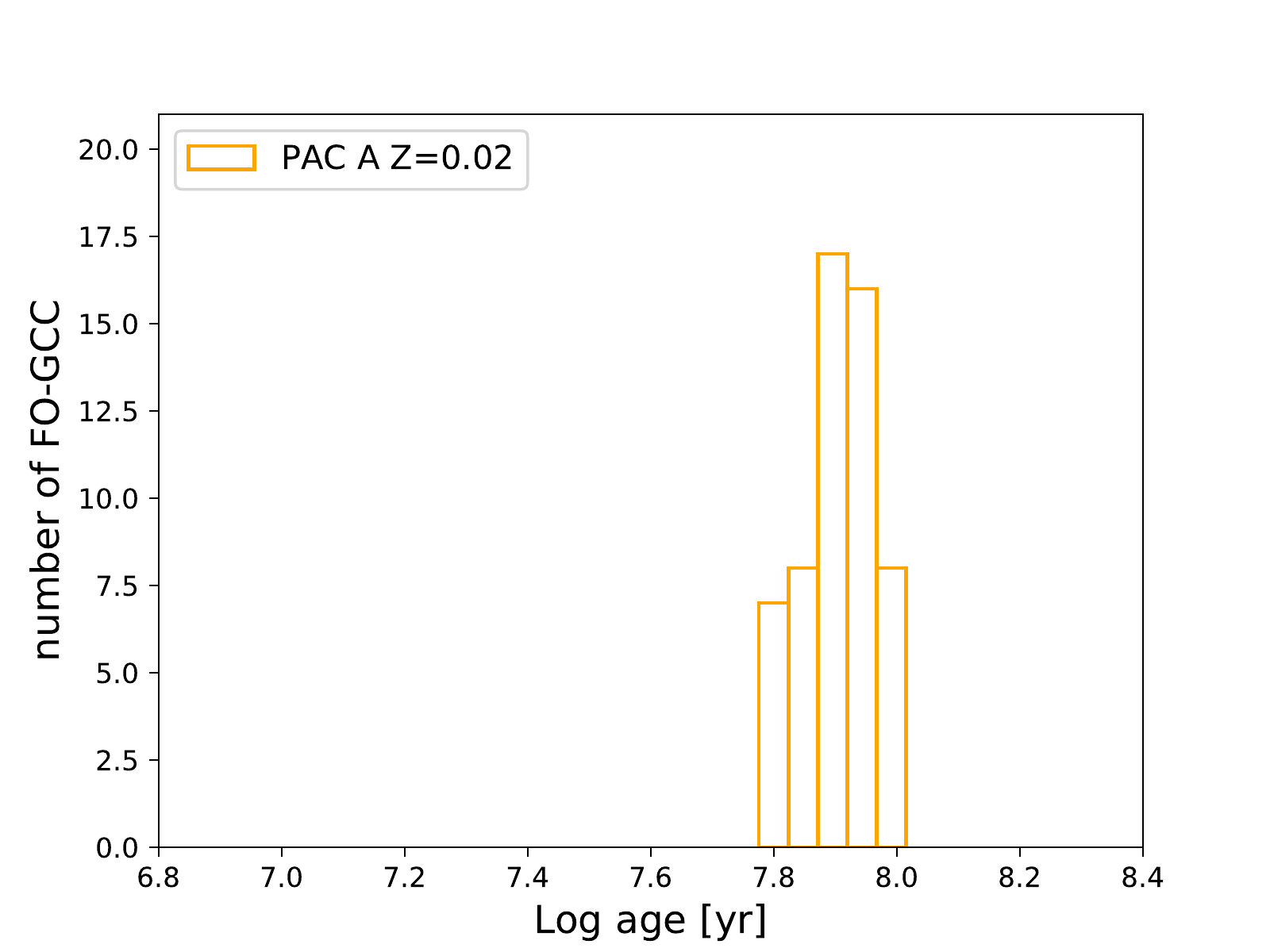}\par 
\end{multicols}
\caption{\label{fig:hist_PAC} The predicted age distribution, derived by applying the PAC relation for $Z$=$0.02$ to the selected sample of F {\sl (left panel)} and FO-mode {\sl (right panel)} CCs, for the assumed cases A and B of the ML relation (see labels).}
\end{figure*}

For the sake of quantifying the impact of the metallicity term on the previous relations, we also plotted, in Fig. \ref{fig:hist_PAC}, the individual age distributions obtained when applying the F and FO-mode PAC relations derived for the solar metallicity case. Regardless of the selected ML relation, the peaks of the retrieved F-mode CC age distributions appear to be shifted towards slightly older ages with respect to the case in which the appropriate CC metallicity is accounted for. For the FO-mode CCs, the age distribution appears quite more peaked than the one obtained from the PACZ relation. 

Assuming the canonical ML relation, the age distribution for fundamental pulsators (see blue bars in the left panels of Figs. \ref{fig:hist_PAZ} and \ref{fig:hist_PACZ}), made using the PAZ and the PACZ relations, peaks around 7.86 and 7.89 Myrs, respectively. The minimum and maximum age values derived from the PAZ relation are 6.99 Myr and 8.20 Myr, respectively. However, the application of the PACZ provides a minimum age of 7.13 Myr and a maximum age of 8.23 Myr. In the noncanonical case (brown bars), the PAZ relation provides a peak age around 7.92 Myr while the PACZ's is around 7.96 Myr. The minimum and maximum age values from the PAZ and PACZ relations are, 7.24 Myr and 8.28 Myr, and 7.32 Myr and 8.24 Myr, respectively. As expected, the entire FO-mode GCC age distribution (see magenta bars in the right panels of Figs. \ref{fig:hist_PAZ} and \ref{fig:hist_PACZ}) is shifted towards older ages with a peak around 7.86 Myr and 7.93 Myr from the PAZ and PACZ relations, respectively; and minimum and maximum age values of 7.61 Myr and 8.05 Myr from the PAZ application and 7.63 Myr and 8.03 Myr from the PACZ one. Since the age errors are smaller in the case of the PACZ, we decided to provide only the ages derived using the PACZ relation. The computed ages of the fundamental pulsators for canonical and noncanonical ML assumptions are partly reported in Tables \ref{age_F_1p5_A_PA} and \ref{age_F_1p5_B_PA}, respectively. The retrieved ages for the canonical first overtone pulsators are partly listed in Table \ref{age_FO_1p5_A_PA}. Complete tables are available in the supplementary material.

For comparison, Fig. \ref{fig:hist_PAC} shows the age distributions derived by applying the PAC relation for the solar metallicity to the same Cepheid sample; that is, neglecting any difference in the individual metal abundances. In particular, the left panel, showing the age distribution for the F-mode pulsators, highlights that the age distributions for canonical (red bars) and noncanonical (green bars) pulsators each peaks around 7.85 Myr and 8.0 Myr, respectively. This is slightly different from the metal-dependent value (7.89 and 7.96 Myr, respectively). The minimum and maximum age values are 7.07 Myr and 8.19 Myr (against 7.13 and 8.23 for the metal-dependent case) for the canonical case, and 7.26 Myr and 8.32 Myr (against 7.32 and 8.24 for the metal-dependent case) for the noncanonical case. The FO-mode pulsator age distribution (right panel) peaks around 7.90 Myr with minimum and maximum age values of 7.77 Myr and 8.01 Myr (against 7.63 and 8.03 for the metal-dependent case).

\subsection{The age distribution maps}

\begin{figure*}
\includegraphics[width=\textwidth]{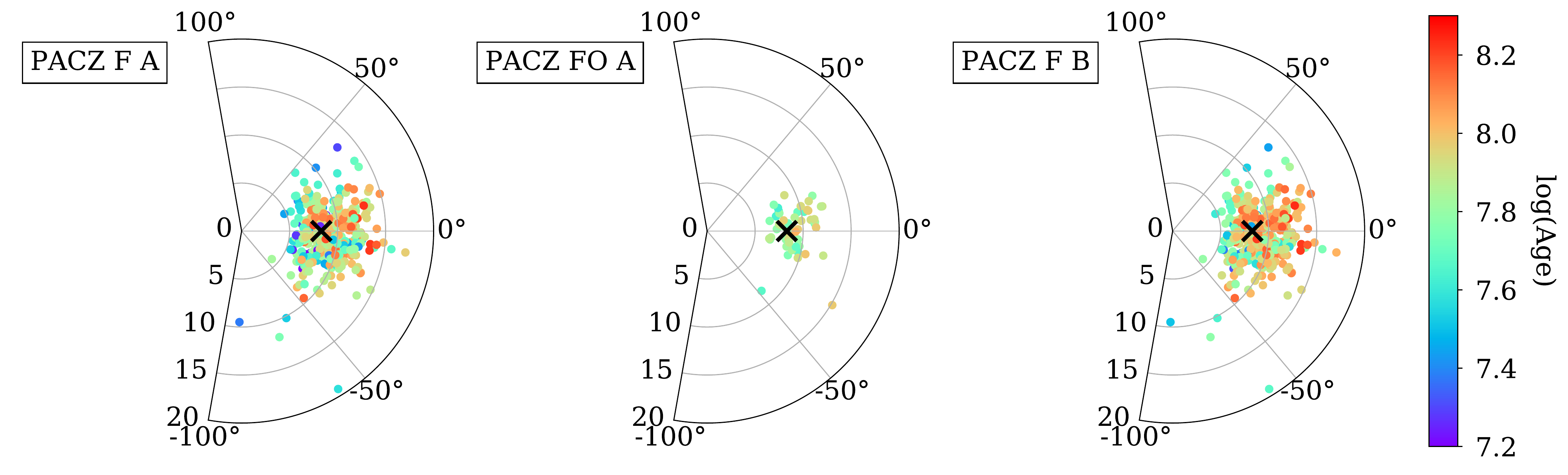}\par
\caption{Distribution of the selected Gaia EDR3 Cepheids on the Galactic plane plotted in polar coordinates for canonical F-mode pulsators (left panel), canonical FO-mode pulsators (middle panel) and noncanonical F-mode pulsators (right panel). The Galactic center is in the middle.  The black ‘X' marks the Sun’s position. In each panel, the colored circles show the predicted individual ages obtained using the PACZ relations, according to the logarithmic color-bar axis}
\label{fig:age_map_PACZ}
\end{figure*}

\begin{figure}
\includegraphics[width=\linewidth]{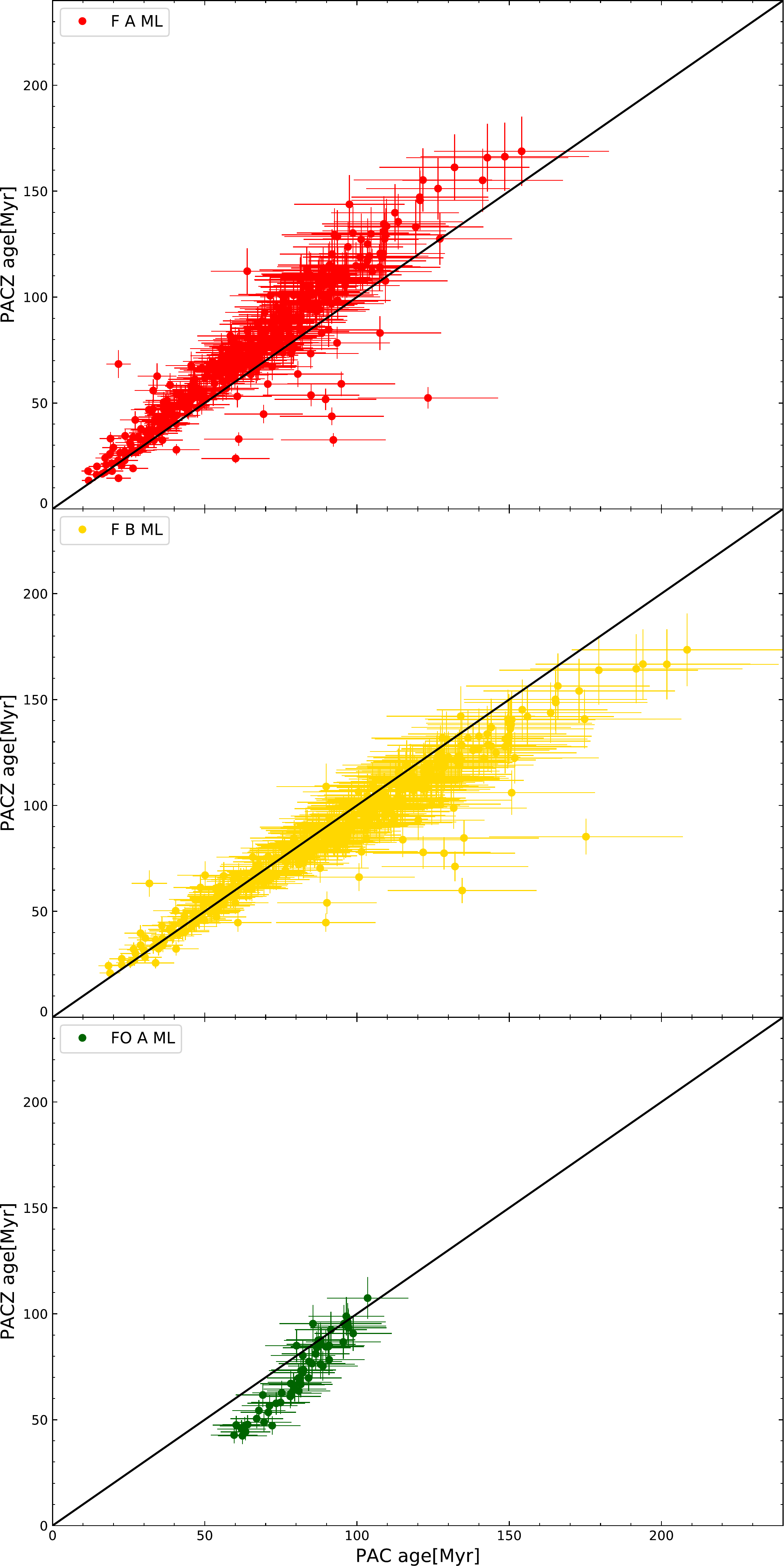}\par
\caption{Comparison between the individual ages obtained by using the
PACZ relation and the PAC relation for $Z$=$0.02$ to the selected sample of canonical F-mode (upper panel), noncanonical F-mode (middle panel) and canonical FO-mode GCCs (bottom panel). The black solid line is the 1:1 line.}
\label{fig:comp_PACZ_PAC_1p5_A}
\end{figure}

Fig. \ref{fig:age_map_PACZ} shows the age distribution map (color-code scale), inferred from the application of the canonical fundamental (left panel), canonical first overtone (middle panel) PACZ relation and the noncanonical fundamental PACZ relation (right panel) to the selected sample of Galactic Cepheids.
To build the figure we adopted the photogeometric distances by \citet{Bailer-Jones2021} based on Gaia EDR3 astrometry and transformed the galactic coordinates into galactocentric cartesian coordinates, assuming that the investigated Cepheids are located on the Galactic disk and that the distance of the Sun from the Galactic centre is 8.3 kpc. The plotted maps show that the selected sample is concentrated at a relatively low Galactocentric distance so that the dependence of the inferred age on this parameter is not evident as in our previous investigation \citepalias{Desomma2020b}. This is due to the limited number of Cepheids in the Gaia EDR3 catalogue with individual metal abundances.
To investigate the effect of the metallicity term in the adopted PAC relation, Fig. \ref{fig:comp_PACZ_PAC_1p5_A} shows the direct comparison between the individual ages inferred from the PACZ relation and the ones obtained from the PAC relation at solar chemical composition, for canonical (upper panel) and noncanonical (middle panel) F-mode and canonical FO-mode (bottom panel) assumptions.

We notice that in the case of the canonical F-mode relations, the inclusion of the metallicity term produces, on average, a slight shift towards older ages, especially moving towards longer periods. On the other hand, for the F-mode noncanonical case, the individual ages predicted by the PACZ relation appear to be, on average, smaller than in the fixed $Z$=$0.02$ case. Finally, in the case of the FO-mode, the difference is significant at shorter periods, with the ages predicted by the PACZ being systematically smaller than the ones obtained from the PAC relation. The comparisons shown in Fig. \ref{fig:comp_PACZ_PAC_1p5_A} suggest that the inclusion of the metallicity term is relevant and it can have a significant effect on the application of PAC relations to constrain the star formation history of Cepheid host galaxies.

\section{Conclusions and Future Developments}

We have presented an updated evolutionary and pulsational scenario for CCs by varying the pulsation mode, the ML relation and the chemical composition. The predicted evolutionary paths and instability strip boundaries were combined to infer updated PA and multi-filter PAC relations, including the first theoretical PAC in the Gaia bands for chemical compositions different from the solar one (see Table \ref{param_puls_models}).

We found that the predicted PA relations, based on the noncanonical scenario, predict older ages than the corresponding canonical PA relations, for any given period and chemical composition; confirming previous results for solar metallicity \citepalias{Desomma2020b}. Moreover, in both the case of the canonical and noncanonical assumptions, as the metal abundance increases, the F-mode PA relation gets flatter. Whereas, no clear trend can be noticed for the FO-mode PA relation apart from a non-negligible variation of its coefficient when varying the chemical composition. A similar dependence on metallicity is found for the PAC coefficients. 
Therefore, since we rely on a theoretical scenario spanning a wide metallicity range, we derived metal dependent PA and PAC relations that we named PAZ and PACZ relations, respectively.

These PAZ and PACZ relations were applied to a subset of Galactic Cepheids from the Gaia EDR3 catalogue with information on the metal abundance. The inferred individual ages show a distribution that confirms that the predicted age is higher for FO-mode pulsators and increases as the ML relation gets brighter. Moreover, the derived age distributions for F and FO-mode pulsators are compared with similar results based on $Z$=$0.02$ models only. As a result we can conclude that, even if the selected sample is concentrated at a relatively low Galactocentric distance the inclusion of the metallicity term produces slightly different individual ages both in the F-mode and FO-mode cases. In particular, in the F-mode canonical case the adoption of the PACZ relation produces, on average, a slight shift towards older ages, especially at longer periods. Whereas, an opposite trend is found in the F-mode noncanonical case and in the FO-mode canonical case.
\clearpage

\begin{table*}
\caption{\label{age_F_1p5_A_PA} Individual ages for the F-mode GCC in our sample obtained by using the canonical PAZ and PACZ relations. The full table is available as supplementary material.}
\centering
\resizebox{\textwidth}{!}{%
\begin{tabular}{cccccccccccccc}
\hline\hline
\textsl{Gaia} EDR3 Source Id & RA[deg] & DEC[deg] & P[d] & [Fe/H] & $\sigma$ [Fe/H] & $G[mag]$ & $G_{BP}$-$G_{RP}$[mag] & E($G_{BP}$-$G_{RP}$)[mag] & $\sigma$E($G_{BP}$-$G_{RP}$)[mag] & $t_{PAZ}$[Myr] & $\sigma$ $t_{PAZ}$[Myr] & $t_{PACZ}$[Myr] & $\sigma$ $t_{PACZ}$[Myr] \\
\hline
(1)&(2)&(3)&(4)&(5)&(6)&(7)&(8)&(9)&(10)&(11)&(12)&(13)&(14)\\
\hline
4757601523650165120 & 83.40631 & -62.48977 & 9.84229 & -0.06 & 0.12 & 3.59 & 0.95 & 0.1 & 0.02 & 44.69 & 8.54 & 54.50 & 5.27 \\
4240272953377646592 & 298.11824 & 1.00562 & 7.17680 & 0.14 & 0.12 & 3.75 & 0.95 & 0.19 & 0.02 & 56.70 & 10.84 & 63.24 & 6.12 \\
3366754155291545344 & 106.02718 & 20.57029 & 10.14860 & -0.11 & 0.12 & 3.54 & 0.93 & 0.06 & 0.03 & 43.72 & 8.36 & 54.87 & 5.31 \\
2200153454733285248 & 337.29288 & 58.41521 & 5.36625 & 0.09 & 0.12 & 3.85 & 0.87 & 0.1 & 0.03 & 71.15 & 13.60 & 81.31 & 7.86 \\
4057701830728920064 & 266.89007 & -27.83083 & 7.01302 & -0.21 & 0.12 & 4.33 & 0.84 & 0.26 & 0.03 & 58.42 & 11.17 & 74.89 & 7.24 \\
4050309195613114624 & 271.25513 & -29.58011 & 7.59498 & 0.08 & 0.12 & 4.59 & 0.96 & 0.15 & 0.02 & 54.37 & 10.39 & 61.03 & 5.90 \\
4514145288240593408 & 284.56145 & 17.36087 & 4.47092 & 0.1 & 0.12 & 5.17 & 0.79 & 0.26 & 0.03 & 81.93 & 15.66 & 97.45 & 9.42 \\
3435571660360952704 & 97.14203 & 30.49298 & 3.72833 & 0.1 & 0.12 & 5.34 & 0.75 & 0.08 & 0.03 & 94.31 & 18.03 & 112.74 & 10.90 \\
1820309639468685824 & 299.00526 & 16.63476 & 8.38234 & 0.14 & 0.12 & 5.46 & 0.93 & 0.16 & 0.02 & 50.27 & 9.61 & 59.18 & 5.72 \\
4096107909387492992 & 275.34576 & -18.86003 & 5.77338 & 0.11 & 0.12 & 5.48 & 0.89 & 0.27 & 0.03 & 67.18 & 12.84 & 76.06 & 7.36 \\
... & ... & ... & ... & ... & ... & ... & ... & ... & ... & ... & ... & ... & ... \\
\hline\hline
\end{tabular}}
\end{table*}

\begin{table*}
\caption{\label{age_F_1p5_B_PA} The same as in Table \ref{age_F_1p5_A_PA} but for noncanonical F-mode models. The full table is available as supplementary material.}
\centering
\resizebox{\textwidth}{!}{%
\begin{tabular}{cccccccccccccc}
\hline\hline
\textsl{Gaia} EDR3 Source Id & RA[deg] & DEC[deg] & P[d] & [Fe/H] & $\sigma$ [Fe/H] & $G[mag]$ & $G_{BP}$-$G_{RP}$[mag] & E($G_{BP}$-$G_{RP}$)[mag] & $\sigma$E($G_{BP}$-$G_{RP}$)[mag] & $t_{PAZ}$[Myr] & $\sigma$ $t_{PAZ}$[Myr] & $t_{PACZ}$[Myr] & $\sigma$ $t_{PACZ}$[Myr] \\
\hline
(1)&(2)&(3)&(4)&(5)&(6)&(7)&(8)&(9)&(10)&(11)&(12)&(13)&(14)\\
\hline
4757601523650165120 & 83.40631 & -62.48977 & 9.84229 & -0.06 & 0.12 & 3.59 & 0.95 & 0.1 & 0.02 & 61.58 & 11.54 & 67.04 & 6.64 \\
4240272953377646592 & 298.11824 & 1.00562 & 7.17680 & 0.14 & 0.12 & 3.75 & 0.95 & 0.19 & 0.02 & 73.13 & 13.71 & 78.43 & 7.77 \\
3366754155291545344 & 106.02718 & 20.57029 & 10.14860 & -0.11 & 0.12 & 3.54 & 0.93 & 0.06 & 0.03 & 60.85 & 11.40 & 66.68 & 6.60 \\
2200153454733285248 & 337.29288 & 58.41521 & 5.36625 & 0.09 & 0.12 & 3.85 & 0.87 & 0.1 & 0.03 & 88.82 & 16.65 & 95.33 & 9.44 \\
4057701830728920064 & 266.89007 & -27.83083 & 7.01302 & -0.21 & 0.12 & 4.33 & 0.84 & 0.26 & 0.03 & 78.34 & 14.68 & 84.96 & 8.41 \\
4050309195613114624 & 271.25513 & -29.58011 & 7.59498 & 0.08 & 0.12 & 4.59 & 0.96 & 0.15 & 0.02 & 71.18 & 13.34 & 75.84 & 7.51 \\
4514145288240593408 & 284.56145 & 17.36087 & 4.47092 & 0.1 & 0.12 & 5.17 & 0.79 & 0.26 & 0.03 & 99.71 & 18.69 & 109.22 & 10.81 \\
3435571660360952704 & 97.14203 & 30.49298 & 3.72833 & 0.1 & 0.12 & 5.34 & 0.75 & 0.08 & 0.03 & 112.04 & 21.00 & 122.79 & 12.16 \\
1820309639468685824 & 299.00526 & 16.63476 & 8.38234 & 0.14 & 0.12 & 5.46 & 0.93 & 0.16 & 0.02 & 66.19 & 12.41 & 73.13 & 7.24 \\
4096107909387492992 & 275.34576 & -18.86003 & 5.77338 & 0.11 & 0.12 & 5.48 & 0.89 & 0.27 & 0.03 & 84.49 & 15.84 & 90.62 & 8.97 \\
... & ... & ... & ... & ... & ... & ... & ... & ... & ... & ... & ... & ... & ... \\
\hline\hline
\end{tabular}}
\end{table*}

\begin{table*}
\caption{\label{age_FO_1p5_A_PA} The same as in Table \ref{age_F_1p5_A_PA} but for canonical FO-mode models. The full table is available as supplementary material.}
\centering
\resizebox{\textwidth}{!}{%
\begin{tabular}{cccccccccccccc}
\hline\hline
\textsl{Gaia} EDR3 Source Id & RA[deg] & DEC[deg] & P[d] & [Fe/H] & $\sigma$ [Fe/H] & $G[mag]$ & $G_{BP}$-$G_{RP}$[mag] & E($G_{BP}$-$G_{RP}$)[mag] & $\sigma$E($G_{BP}$-$G_{RP}$)[mag] & $t_{PAZ}$[Myr] & $\sigma$ $t_{PAZ}$[Myr] & $t_{PACZ}$[Myr] & $\sigma$ $t_{PACZ}$[Myr] \\
\hline
(1)&(2)&(3)&(4)&(5)&(6)&(7)&(8)&(9)&(10)&(11)&(12)&(13)&(14)\\
\hline
6058439910929477120 & 187.91793 & -59.42394 & 3.34265 & -0.11 & 0.12 & 5.27 & 0.70 & 0.14 & 0.03 & 70.04 & 6.79 & 67.07 & 6.18 \\
5506374096132016512 & 114.57584 & -48.60141 & 5.69453 & -0.11 & 0.12 & 5.47 & 0.76 & 0.14 & 0.02 & 46.67 & 4.52 & 45.60 & 4.20 \\
5519380081746387328 & 122.99983 & -46.64432 & 4.22727 & 0.09 & 0.12 & 5.59 & 0.74 & 0.05 & 0.03 & 56.47 & 5.47 & 53.50 & 4.93 \\
1853025642297186688 & 316.62604 & 31.18463 & 2.49921 & 0.16 & 0.12 & 5.65 & 0.67 & 0.09 & 0.03 & 83.22 & 8.06 & 75.29 & 6.93 \\
1964855904803120640 & 319.84243 & 38.23746 & 3.33282 & 0.09 & 0.12 & 5.72 & 0.70 & 0.03 & 0.0 & 67.69 & 6.56 & 62.47 & 5.75 \\
541716257882226560 & 42.99483 & 68.88847 & 1.95151 & 0.12 & 0.12 & 5.73 & 0.63 & 0.37 & 0.04 & 101.22 & 9.81 & 90.72 & 8.36 \\
3409635486731094400 & 69.31155 & 18.54301 & 3.14804 & 0.15 & 0.12 & 6.27 & 0.76 & 0.41 & 0.04 & 69.93 & 6.78 & 67.70 & 6.24 \\
5960623272099513856 & 264.40651 & -40.81355 & 3.38033 & 0.18 & 0.12 & 7.07 & 0.72 & 0.34 & 0.04 & 65.87 & 6.38 & 61.00 & 5.62 \\
5848500161483878400 & 222.62620 & -67.49763 & 3.06522 & 0.14 & 0.12 & 7.14 & 0.71 & 0.5 & 0.04 & 71.49 & 6.93 & 66.67 & 6.14 \\
5877460679352962048 & 221.67487 & -61.46196 & 2.39810 & -0.01 & 0.12 & 7.33 & 0.70 & 0.35 & 0.06 & 88.58 & 8.58 & 84.94 & 7.82 \\
... & ... & ... & ... & ... & ... & ... & ... & ... & ... & ... & ... & ... & ... \\
\hline\hline
\end{tabular}}
\end{table*}

\section*{Acknowledgements}
The authors thank the anonymous Referee for the pertinent and useful comments that significantly improved the content and readability of the manuscript.
We acknowledge the financial support from ASI-Gaia (“Missione Gaia Partecipazione italiana al DPAC – Operazioni e Attività di Analisi dati”), and the project Mainstream INAF "Stellar evolution and asteroseismology in the context of PLATO space mission" (PI: S.Cassisi). GDS thanks the support from Istituto Nazionale di Fisica Nucleare (INFN), Naples section-Specific Initiative Moonlight2.

Data availability statement:The data underlying this article are available in the article and in its online supplementary material.



\bibliographystyle{mnras}
\bibliography{desomma_main} 


\bsp	
\label{lastpage}
\end{document}